\documentclass[twocolumn]{aastex631}
\usepackage{comment}
\usepackage{booktabs}

\begin{document}

\title{A High-Resolution NUV Transmission Spectrum of KELT-9b: Mg II and Fe II Escaping from the Hottest Known Giant Planet}

\author{Austin Baldwin}
\affiliation{Department of Physics,
Utah Valley University,
800 W University Pkwy,
Orem, UT 84058, USA}

\author[0000-0003-3667-8633]{Joshua D. Lothringer}
\affiliation{Space Telescope Science Institute,
3700 Charles St.,
Baltimore, MD 21218}

\author[0000-0002-2248-3838]{Leonardo A. dos Santos}
\affiliation{Space Telescope Science Institute,
3700 Charles St.,
Baltimore, MD 21218}

\author[0000-0001-6050-7645]{David K. Sing}
\affiliation{Department of Earth and Planetary Sciences, Johns Hopkins University, Baltimore, MD}
\affiliation{William H.\ Miller III Department of Physics \& Astronomy,
Johns Hopkins University, 3400 N Charles St, Baltimore, MD 21218, USA}

\author[0000-0001-6050-7645]{Zafar Rustamkulov}
\affiliation{Department of Earth and Planetary Sciences, Johns Hopkins University, Baltimore, MD}

\author[0000-0002-6500-3574]{Nikolay K. Nikolov}
\affiliation{Space Telescope Science Institute,
3700 Charles St.,
Baltimore, MD 21218}

\author[0000-0003-3667-8633]{Jeff Valenti}
\affiliation{Space Telescope Science Institute,
3700 Charles St.,
Baltimore, MD 21218}

\author[0000-0003-4328-3867]{Hannah R. Wakeford}
\affiliation{School of Physics, University of Bristol, H.H. Wills Physics Laboratory, Bristol, UK}

\begin{abstract}
We present high-resolution NUV observations from Hubble Space Telescope's (HST) Space Telescope Imaging Spectrograph (STIS) data for the hottest known gas planet, KELT-9b. Observations were collected with STIS/E230M (2300-3000 $\mathrm{\AA}$, R$\sim$ 30,000) and we de-correlate systematic effects from the telescope using jitter detrending. We show the clear presence of the Mg II doublet at 2800 $\mathrm{\AA}$ and Fe II at 2600 $\mathrm{\AA}$ in KELT-9b. The Mg II is measured above the planet's Roche transit radius, indicating it is escaping. We fit 1D NLTE atmospheric escape models to these features, demonstrating a significant loss of mass in KELT-9b's atmosphere ($\dot{M} \approx 10^{12} $ g/s); we also find a remarkably high line-broadening corresponding to a velocity of about $50-75$ km/s, and a net blueshift of the Mg II doublet greater than 30 km/s. Future 3D MHD modeling of the spectrum and gas kinematics is likely needed to explain these observations. We interpret these results in the context of the Mg II ``Cosmic Shoreline" and show that the detection of escaping Mg II in KELT-9b and the non-detection in WASP-178b are consistent with the hypothesis that stars hotter than $T_{\mathrm{eff}} \sim$ 8250~K have relatively low levels of XUV radiation due to the lack of a chromosphere. Therefore planets around such early-type stars experience a different degree of atmospheric escape. This result highlights the importance of XUV irradiation in driving atmospheric escape inside and outside the Solar System.


\end{abstract}

\received{May 19, 2025}

\revised{November 10, 2025}

\revised{December 5, 2025}

\accepted{December 21, 2025}

\submitjournal{AAS Journals}

\keywords{Hot Jupiters (753), Exoplanet Atmospheres (487), Exoplanet atmospheric dynamics (2307), Exoplanet atmospheric evolution (2308), Transmission spectroscopy (2133), Ultraviolet spectroscopy (2284)}

\section{Introduction} \label{sec:intro}

Of all known planets, KELT-9b undergoes the most extreme irradiation \citep{gaudi:2017}. KELT-9b receives more bolometric energy than any other planet, as demonstrated by its nearly 4050~K equilibrium temperature -- almost 1000~K hotter than the next hottest planet, TOI-2109 \citep{wong:2021:2109}. Put another way, KELT-9b experiences the same amount of irradiation every second as twenty 10$^{33}$ erg super-flares hitting TRAPPIST-1b \citep{glazier:2020}. Just as significant, KELT-9b also orbits the hottest known exoplanet host star, which at $T_{\mathrm{eff}}$ = 10,170 $\pm$ 450~K outputs almost exactly half of its energy at ultraviolet wavelengths.

This extreme irradiation drives significant atmospheric mass loss. Initial theoretical estimates placed KELT-9b's mass loss rate between 10$^{10}$ and 10$^{13}$ g/s \citep{gaudi:2017}. More recently modeling estimates escape rates of 10$^{11}$-10$^{12}$ g/s \citep{garcia-munoz:2019} and 10$^{12}$-10$^{13}$ g/s \citep{shaikhislamov:2025}. \cite{cauley:2019} used observations of H-Balmer and Mg I to infer 10$^{12}$ - 10$^{12.5}$ g/s. Subsequent observations of H$\alpha$ and H$\beta$ measured a mass-loss rate of $10^{12.8 \pm 0.3}$ g/s and a thermospheric temperature of $13200^{+800}_{-720}$~K when LTE is assumed \citep{wyttenbach:2020}. Self-consistent NLTE modeling matching the observed H$\alpha$ and H$\beta$ profiles suggests a maximum temperature of about 8500~K \cite{fossati:2021}. NLTE modeling of escaping oxygen lines was used to measure an escape rate of 10$^{11}$-10$^{12}$ g/s \citep{borsa:2022}. 

Recently, \cite{Lowson:2023} found that the shape of the H$\alpha$ light-curve deviated from a normal transit shape, perhaps suggestive of a cometary tail of escaping gas, reminiscent of the asymmetric shapes seen in \cite{cauley:2019}. \cite{stangret:2024} and \cite{DArpa24} have shown evidence for significant turbulent line broadening of up to $\sim$10 km/s through observations of individual metal lines. The Fe II lines were also shown to be significantly blueshifted in \cite{stangret:2024}. \cite{egan:2024} searched for the NUV Mg II doublet with the CUTE CubeSat \citep{france:2023}, and while a general increase in absorption was found in the NUV, the observations were not sensitive enough to constrain the presence of the Mg II doublet at 2800\,$\rm \AA$.

In addition to the mass-loss rate, KELT-9b is also known to host an extensive inventory of detected elements. In total, H, Ca, Ca II, Cr, Cr II, Fe, Fe II, Mg, Mg II, Na, Na II, Ni, O, Sc, Sc II, Si, Sr II, Tb II, Ti, Ti II, and Y II have all been detected with high-resolution spectroscopy \citep{yan:2018,hoeijmakers:2018a,hoeijmakers:2019,yan:2019,cauley:2019, wyttenbach:2020,pino:2020,turner:2020,kasper:2021,borsa:2022,PaiAsnodkar:2022,bello-arufe:2022,Langeveld22,Sanchez-Lopez22,Lowson:2023,Ridden-Harper23,Borsato23,DArpa24}.

Fortunately, KELT-9b's NUV-bright host star provides a unique lens through which we can understand the planet. Ultraviolet transmission spectroscopy of exoplanets is usually limited by the sheer number of photons we can detect coming from the host star. For planets like KELT-9b, however, we can obtain high-SNR constraints on the UV transit spectra of the planet, where species like Fe, Mg, and SiO absorb \citep{sharp:2007,lothringer:2020b}. Such observations have revealed Mg II and Fe II escaping from ultra-hot Jupiter WASP-121b \citep{sing:2019} and strong absorption from some combination of Fe, Mg, and SiO in ultra-hot Juipter WASP-178b \citep{lothringer:2022}.

The escaping Mg II and Fe II from WASP-121b was found with the high-resolution E230M echelle grating on the Space Telescope Imaging Spectrograph (STIS) aboard the \textit{Hubble Space Telescope}. At a resolving power of R$\sim$30,000, E230M is second only to STIS's E230H in terms of NUV spectral resolution of any available instrument, enabling both components of the Mg II doublet to be resolved with velocites of $\pm\sim$10~km/s. 

While the canonical hot Jupiters HD 189733b and HD 209458b have been both observed with STIS/E230M, their spectra have resulted in inconclusive or contested detections \citep{vidal-madjar:2013,cubillos:2020,cubillos:2023}. HD 189733b has also been observed with HST/COS, finding evidence for escaping C \citep{dossantos:2023}. Hot Jupiter WASP-79b is the only other planet with published observations with STIS/E230M, and while a rise in the transit depth of the NUV continuum was measured, there was no clear evidence of hydrodynamical escape \citep{gressier:2023}. Lastly, similar gratings with HST/COS have been used to observe WASP-12b, finding some evidence for enhanced transit depths near metal absorption lines \citep{fossati:2010} potentially due to a cloud of circumstellar gas \citep{fossati:2013} potentially causing an early ingress \citep[][but see \citealt{turner:2016}]{vidotto:2010}.

Whether or not atmospheric escape is occurring for a given planet or planet population may be viewed from the perspective of the ``Cosmic Shoreline" \citep{zahnle:2017}. The Cosmic Shoreline is a hypothesis to explain why some Solar System bodies retain an atmosphere: Objects with high escape velocity, $V_\mathrm{esc}$, and low incident (or accumulated) high-energy flux, $F_\mathrm{XUV}$, tend to have thick atmospheres, while objects with low $V_\mathrm{esc}$ and high incident $F_\mathrm{XUV}$ tend to be airless. This relationship is suggestive of atmospheric escape determining the fate of planetary atmospheres. One can use the same properties, $F_\mathrm{XUV}$ and $V_\mathrm{esc}$, to sort which giant planets might be expected to exhibit ongoing atmospheric mass-loss. The degree to which such a relationship holds for various escape indicators like Ly-$\alpha$, metastable He, or Mg II is not well known, but charting such a Cosmic Shoreline for giant planets may inform searches for atmospheres around smaller planets, for which the Cosmic Shoreline is a guiding framework \citep{redfield:2024,lustig_yaeger:2025:prop}.

Here, we present the high-resolution NUV transmission spectrum of KELT-9b from HST/STIS/E230M. In Section~\ref{sec:methods}, we descibe the observations and the methods used to reduce them. We also explain our modeling set up in Section~\ref{sec:methods:models}. We show the results in Section~\ref{sec:results} and discuss them in Section~\ref{sec:disc} before finally concluding in Section~\ref{sec:conc}.

\begin{figure}
    \centering
    \includegraphics[width=1.0\linewidth]{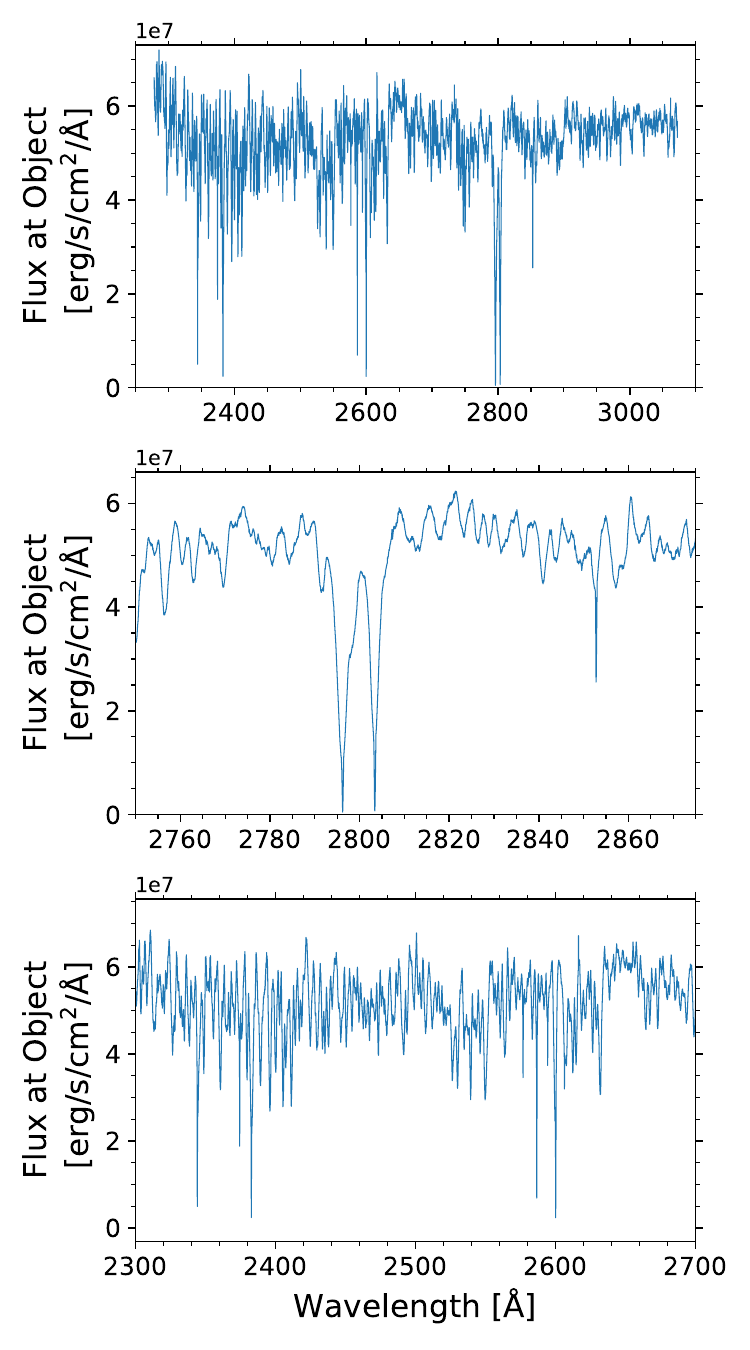}
    \caption{Top: Program-level co-added HST/STIS/E230M spectrum of host star KELT-9. Middle: Same as above, but zoomed in on the Mg II doublet at 2796.35 and 2803.53~$\mathrm{\AA}$ and an Mg I line at 2853.00~$\mathrm{\AA}$. Bottom: Same as above, but zoomed in on Fe II lines at 2344.25, 2382.80, 2586.69, and 2600.21~$\mathrm{\AA}$.}
    \label{fig:stellar_spec}
\end{figure}

\section{Data Analysis} \label{sec:methods}

We observed two transits of KELT-9b with the Hubble Space Telescope (HST) using STIS/E230M (R$\sim$30,000) from about 2,300 to 3,100 $\mathrm{\AA}$ (central wavelength 2,707$\mathrm{\AA}$) on September 22, 2021 and Jan, 19, 2022. Five HST orbits were spent observing the KELT-9b system during each of two transits. Each visit was composed of seven exposures on the first orbit and 10 on the subsequent four, for a total of 47 spectra per transit. As shown in Figure \ref{kelt9b_detrend}, the transit occurred on the first visit during HST's second to fourth orbits, leaving the first and fifth orbits as baseline stellar flux; on the second visit, only the second and third orbits fell during a transit. Figure~\ref{fig:stellar_spec} shows the program-level co-added NUV surface-flux spectrum of the star, KELT-9 using the Hubble Advanced Spectral Product, scaled to the distance and radius of the star \citep[HASP,][]{debes:2024}.

The first orbit and the early exposures of each subsequent orbit have uncharacteristically large fluxes due to other unaccounted for systematics. Observing the spectra of these earlier orbits reveals that this data also exhibits large amounts of random noise. Because of this, we choose to discard the first exposure of each orbit, as in \cite{sing:2019}, though we retain the rest of the first orbit in fitting the data.

 We created a custom Python pipeline to import the STIS x1d.fits spectrum files from the latest version of \texttt{calstis}. The \texttt{calstis} pipeline performs basic reduction and calibration of the data, including 1D spectral extraction and a wavelength correction to a heliocentric reference frame, correcting for HST's orbital motion. We also made use of the orbit-by-orbit .jit instrumental jitter files. HST captures telemetry data alongside its science data and stores them in the jitter files: where the telescope is pointing, positions of guide stars in the detector, the current time, certain binary flags indicating various statuses of the data, the direction from HST to Earth and the Sun, among others. We use these measurements as detrending vectors to account for the significant visit-long and orbit-long systematics seen in the raw data.
 
\begin{figure}[ht!]
\includegraphics[width=\linewidth]{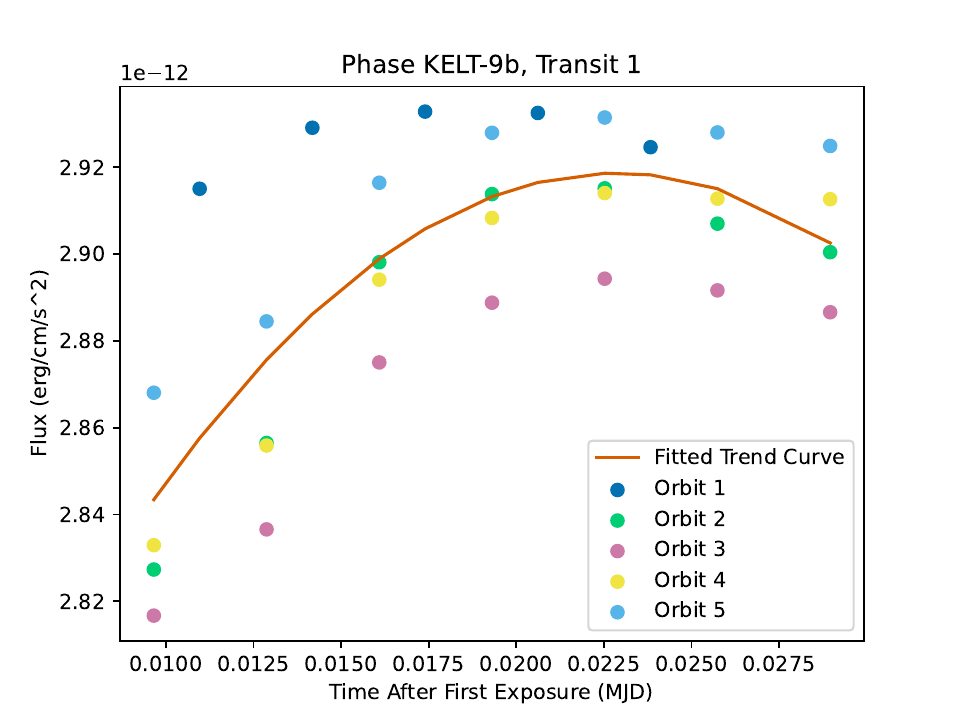}
\caption{The initial second-degree systematics vector fitting for the first transit, with respect to HST's orbital phase, represented as the time since the first exposure in the orbit.
\label{fig:jitterfit}}
\end{figure}

\subsection{Detrending}

We extracted each spectrum's white-light flux from the x1d.fits files and first fit a polynomial trend curve for each systematic variable from the .jit instrument data. The polynomial coefficients were used as initial guesses for a total jointly-fit systematics model later used in our final systematics model fits with \texttt{scipy.optimize.curve\_fit}:

\begin{equation}\label{eq:one}
 S(x) = \Sigma a_n s_n(x)   
\end{equation}

\noindent where $a_n$ are the raw systematics vectors, $s_n$ are polynomials multiplied against the systematics vectors, and $S(x)$ is the total systematics model obtained by summing over all individual systematics trends.

To determine which systematics to detrend from our data, we compared the data and tried a suitable degree for the polynomial trend to fit the variable. If we observed visible curvature between the variable and flux, as in \ref{fig:jitterfit}, we assigned a 2nd degree quadratic fit function; otherwise we assign a 1st degree linear fit function. There was one exception to this method: \texttt{Time} exhibits such a higher-order curve, but since fitting a non-linear trend to the time axis can dilute the fitted transit depth we instead assigned it a linear trend function. 

To further quantify these trends, we used the square of the Pearson linear correlation coefficient, $R^2$, calculating the single-variable correlation between the systematics data and the white-light observations. We used these scores to aid in selecting which variables would be fit in our final systematics model in order to avoid overfitting - those with $R^2 > 0.1$, where less than 10\% of the variance in flux can be explained by the trend line. This produces some inconsistencies in the selection of parameters, where $R^2$ scores vary significantly from visit 1 to visit 2. We list some of these $R^2$ values in Table~\ref{table:pearson_coefficients}. In the end, we used a linear trend in time and second-order polynomials fit to Phase, V2\_roll, and V3\_roll to both visits, with LimbAng and Mag\_V2 fit for Visit 1 and Longitude fit for Visit 2.

\begin{figure*}
\gridline{\fig{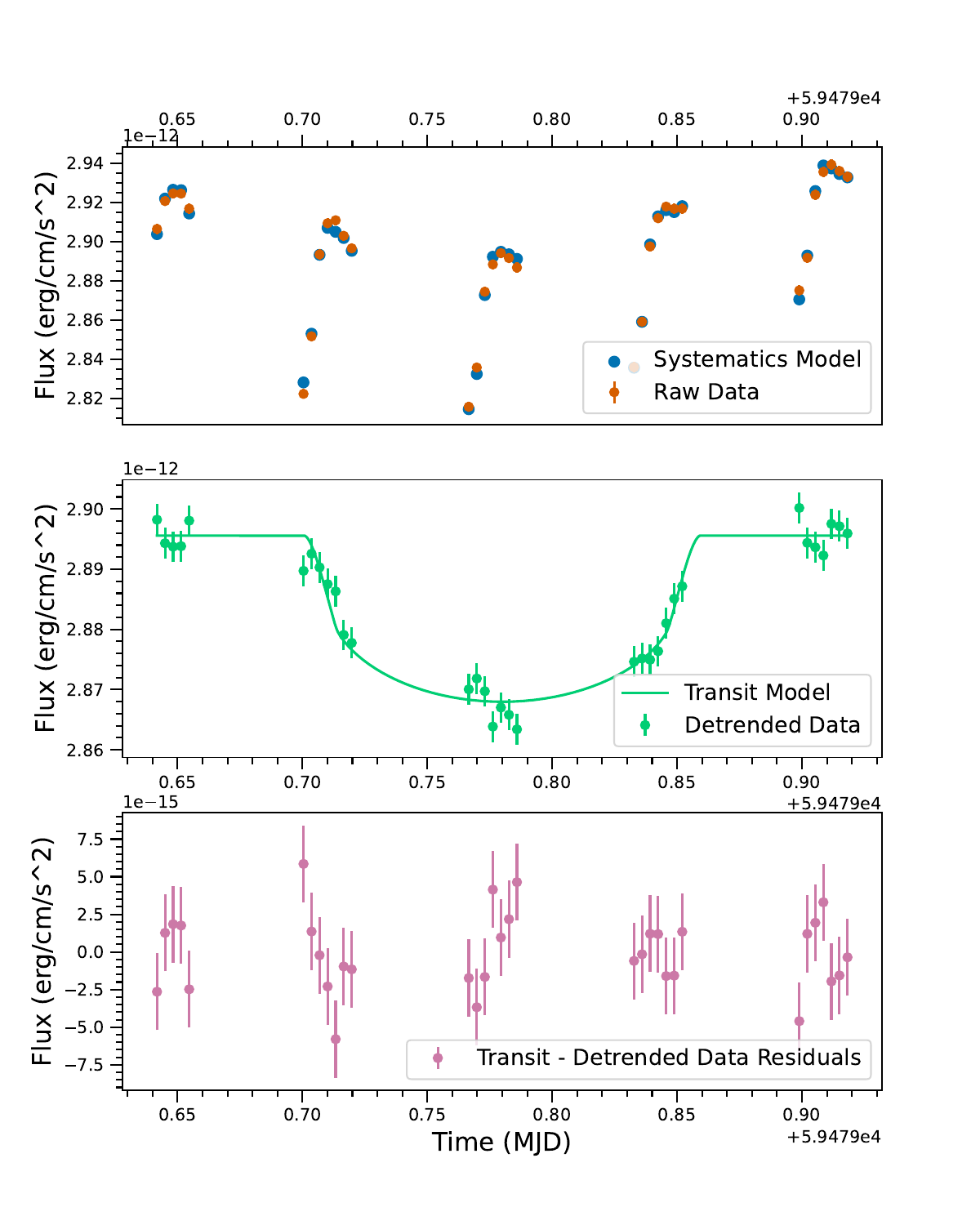}{0.52\textwidth}{Visit 1}
          \fig{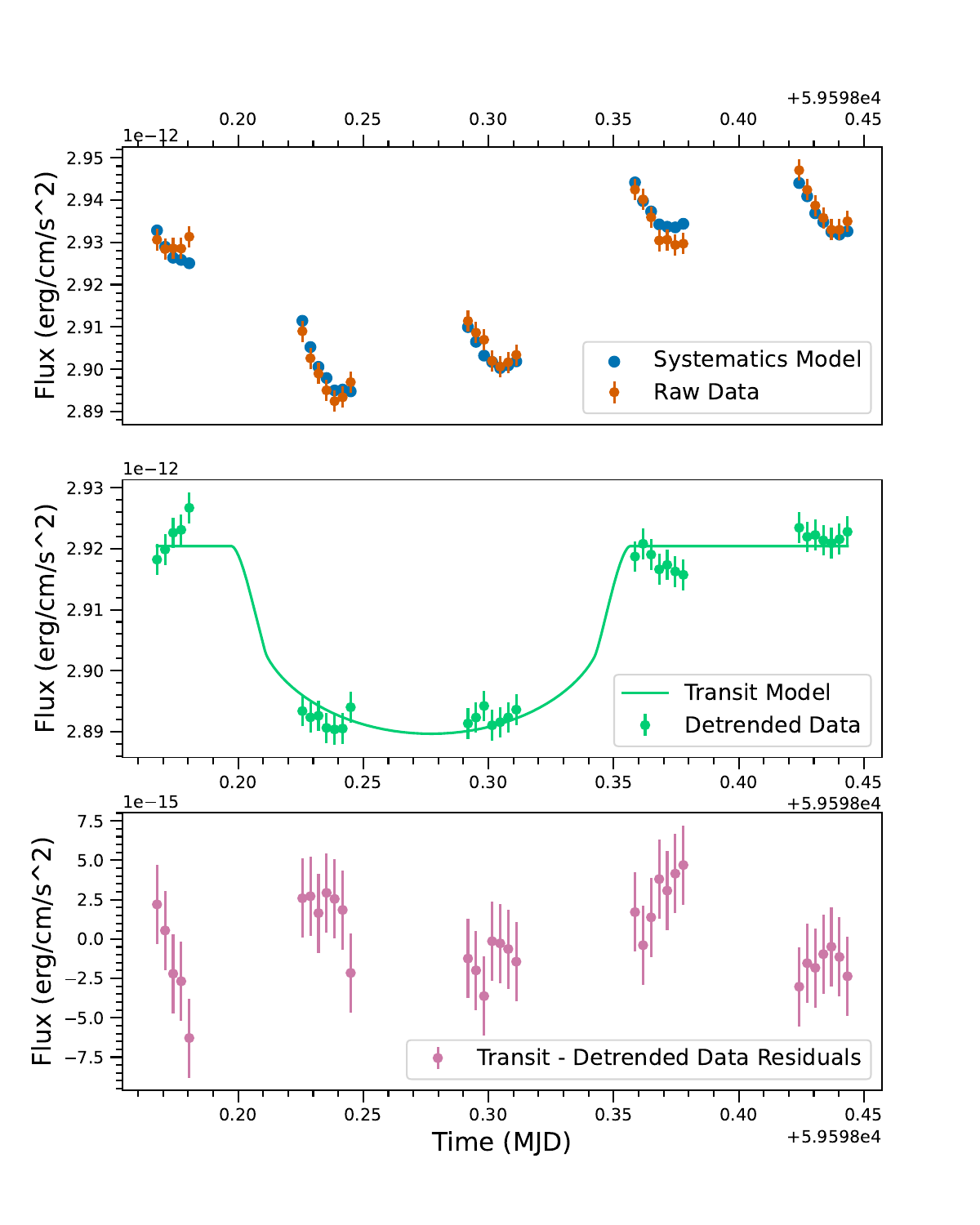}{0.52\textwidth}{Visit 2}}
\caption{Top: The raw STIS/E230M white-light curve for KELT-9b for Visit 1 (left) and Visit 2 (right) with the fitted systematics model. Middle: The systematics-subtracted light curve, fitted to the detrended data. Bottom: The residuals between the detrended data and the transit model. For the first visit, we find a standard deviation about 0.98\% of the data, while for the second visit, we find residual scatter of about 0.38\%.}
\label{kelt9b_detrend}
\end{figure*}

We found that despite its low $R^2$ score on visit 2, including \texttt{Phase} was necessary to create a good detrending model. Additionally, those variables with good fits did not necessarily have important contributions to the model individually.

\begin{deluxetable}{l|r|r}
    \tablecaption{$R^2$ Pearson correlation coefficients for selected trend models \label{table:pearson_coefficients}}
    \tablehead{\colhead{Parameter} & \colhead{Visit 1} & \colhead{Visit 2}}
    \startdata
    Phase \phantom{hellolotsofspace} & \phantom{hello} 0.302 & \phantom{hello} 0.050\\
    V2\_roll & 0.347 & 0.374\\
    V3\_roll & 0.396 & 0.419\\
    Time & 0.028 & 0.275\\
    Longitude & 0.035 & 0.135\\
    LimbAng & 0.323 & 0.00002\\
    Mag\_V2 & 0.72 & 0.013\\
    \enddata
\end{deluxetable}

After generating our initial out-of-transit fit to the systematics model, we used the \texttt{batman} transit model \citep{kreidberg:2015:batman} to generate a model $f(t)$, created by combining a transit model $T(t, \theta)$, stellar baseline flux $F_0$, and the systematics model $S(x)$ from Equation~\ref{eq:one}, in a similar manner to \cite{sing:2019}:

\begin{equation}\label{eq:two}
    f(t) = T(t, t_0, r_p) \times F_0 \times S(x),
\end{equation}

\noindent where $t$ is time, $t_0$ is the transit center, and $r_p$ is the planet radius.

We used \texttt{scipy.optimize.curve\_fit} to fit the unknown transit parameters: $t_0$, $r_p$, and the systematics coefficients $a_n$ simultaneously. The remaining, known transit parameters (period, inclination, and semi-major axis) were sourced from \cite{gaudi:2017}, which remains the most precise to date according to the NASA Exoplanet Archive \citep{christiansen:2025}. The star's quadratic limb darkening parameters were calculated using a custom PHOENIX stellar model \citep{hauschildt:1999} with fundamental parameters from \cite{gaudi:2017}. This produced a model for the white light curve shown in Figure~\ref{kelt9b_detrend}, with best-fit $t_0$ and $r_p$ parameters listed in Table~\ref{table:transit_best_fits}. 

\begin{figure*}
\gridline{\fig{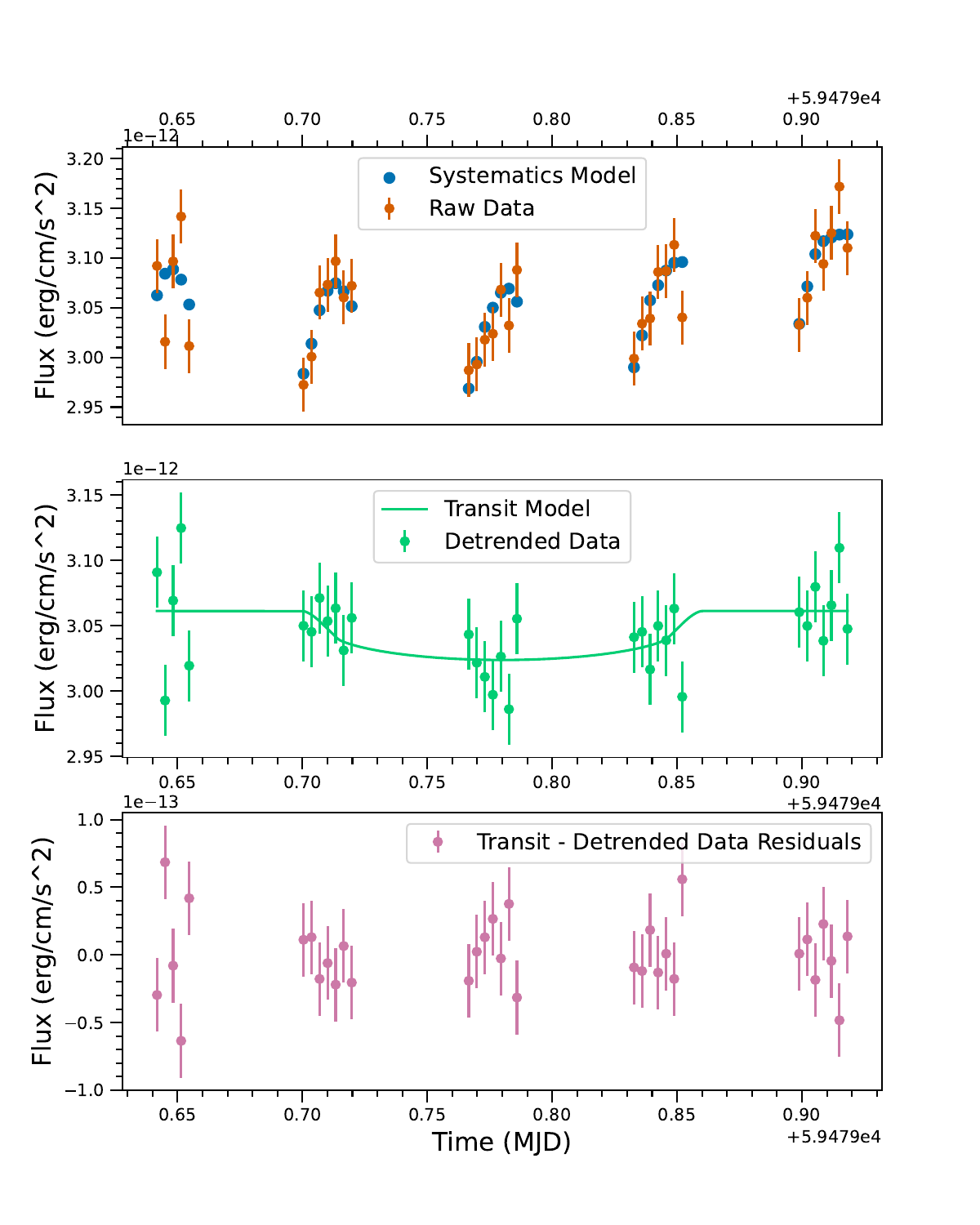}{0.52\textwidth}{Outside Mg II (2700.2 $\mathrm{\AA}$)}
\fig{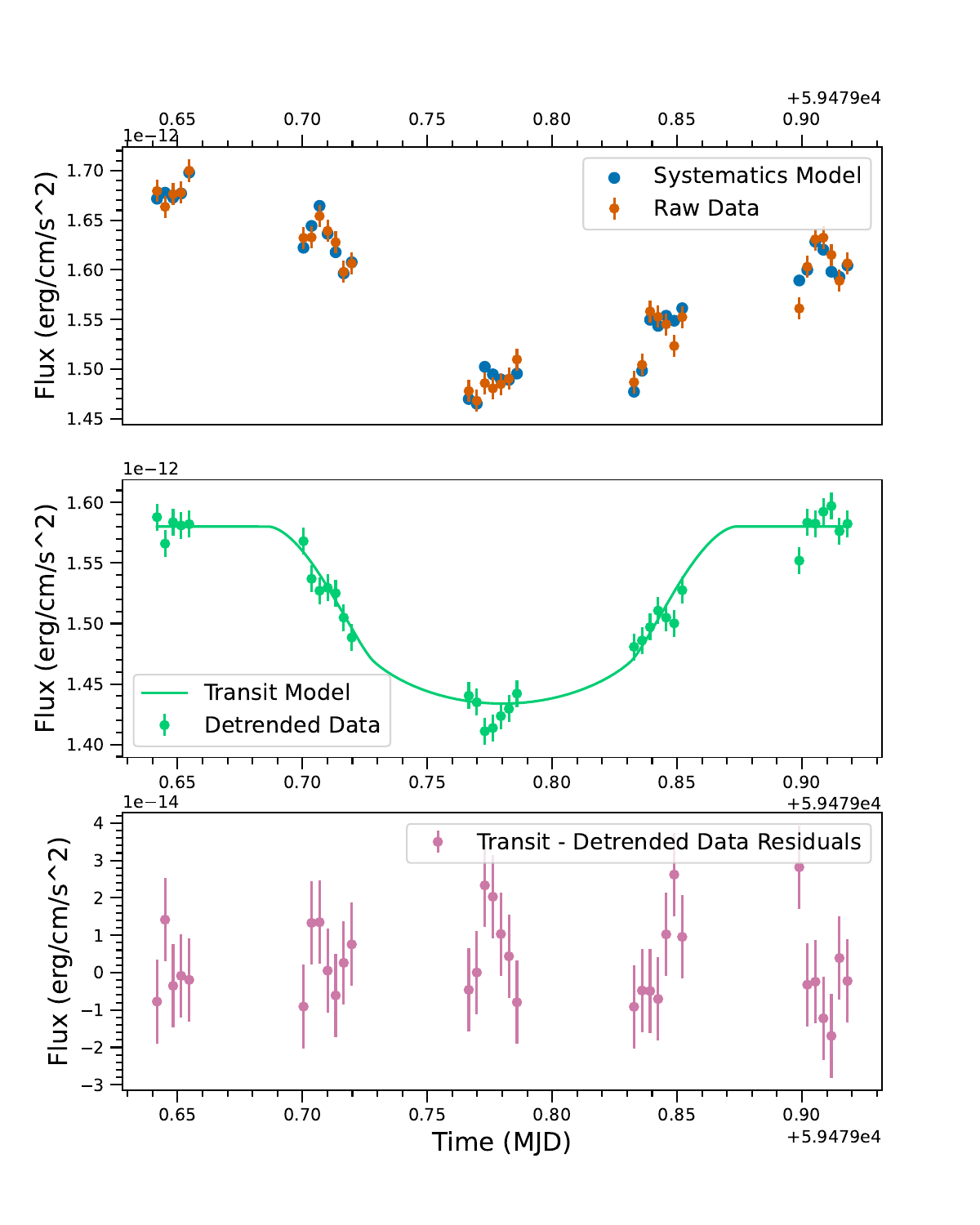}{0.52\textwidth}{Inside Mg II (2802.4 $\mathrm{\AA}$)}}
\caption{Same as Figure~\ref{kelt9b_detrend}, but for two 1-$\mathrm{\AA}$ wide bins from Visit 1, centered at 2700.2 $\mathrm{\AA}$ (left) and 2802.4 $\mathrm{\AA}$, respectively outside and inside the Mg II planetary absorption. Observe that on the second plot, the overall flux is nearly half as bright due to absorption from the host star.}
\label{kelt9b_spec_LC_v1}
\end{figure*}

\begin{figure*}
\gridline{\fig{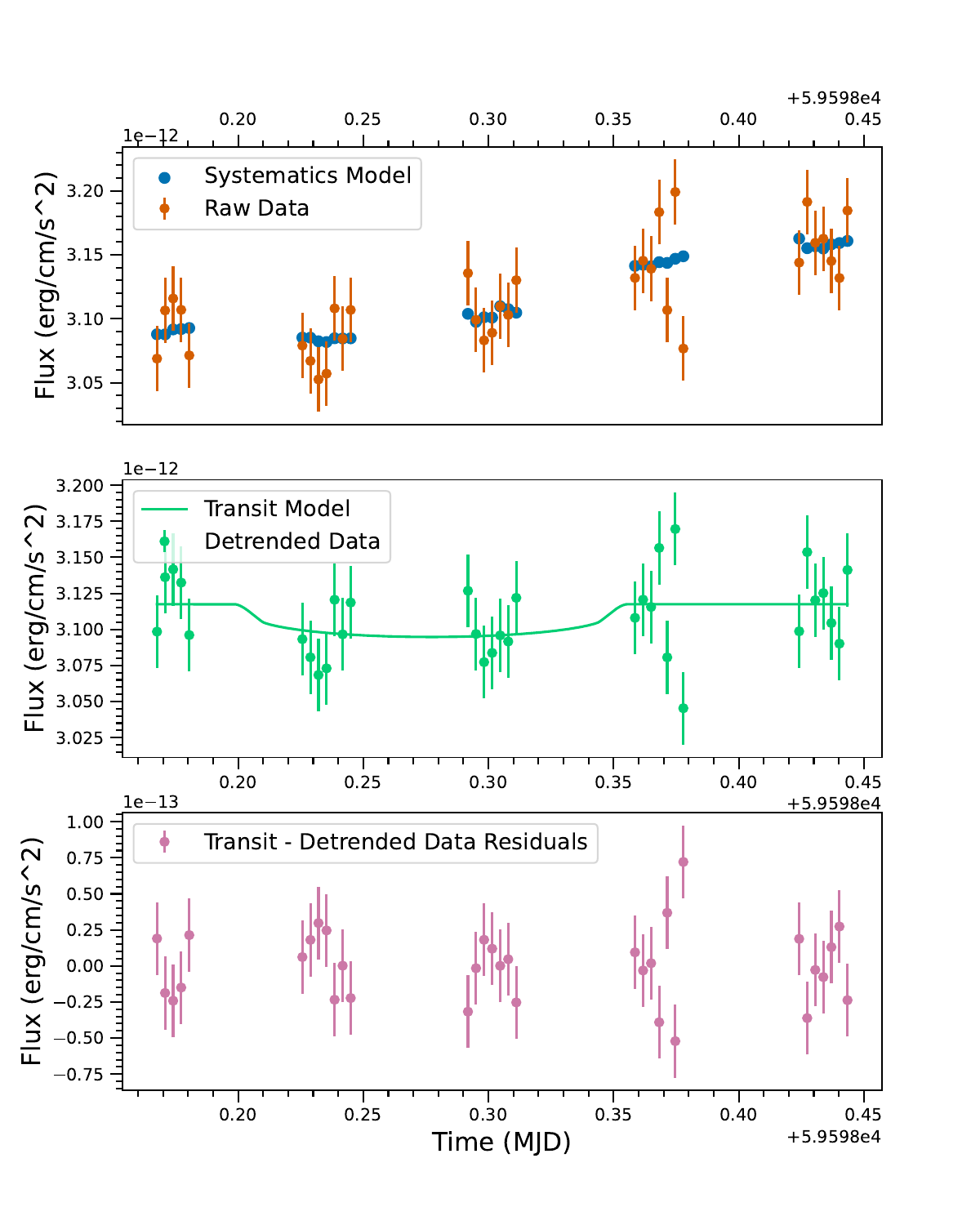}{0.52\textwidth}{Outside Mg II (2700.2 $\mathrm{\AA}$)}
\fig{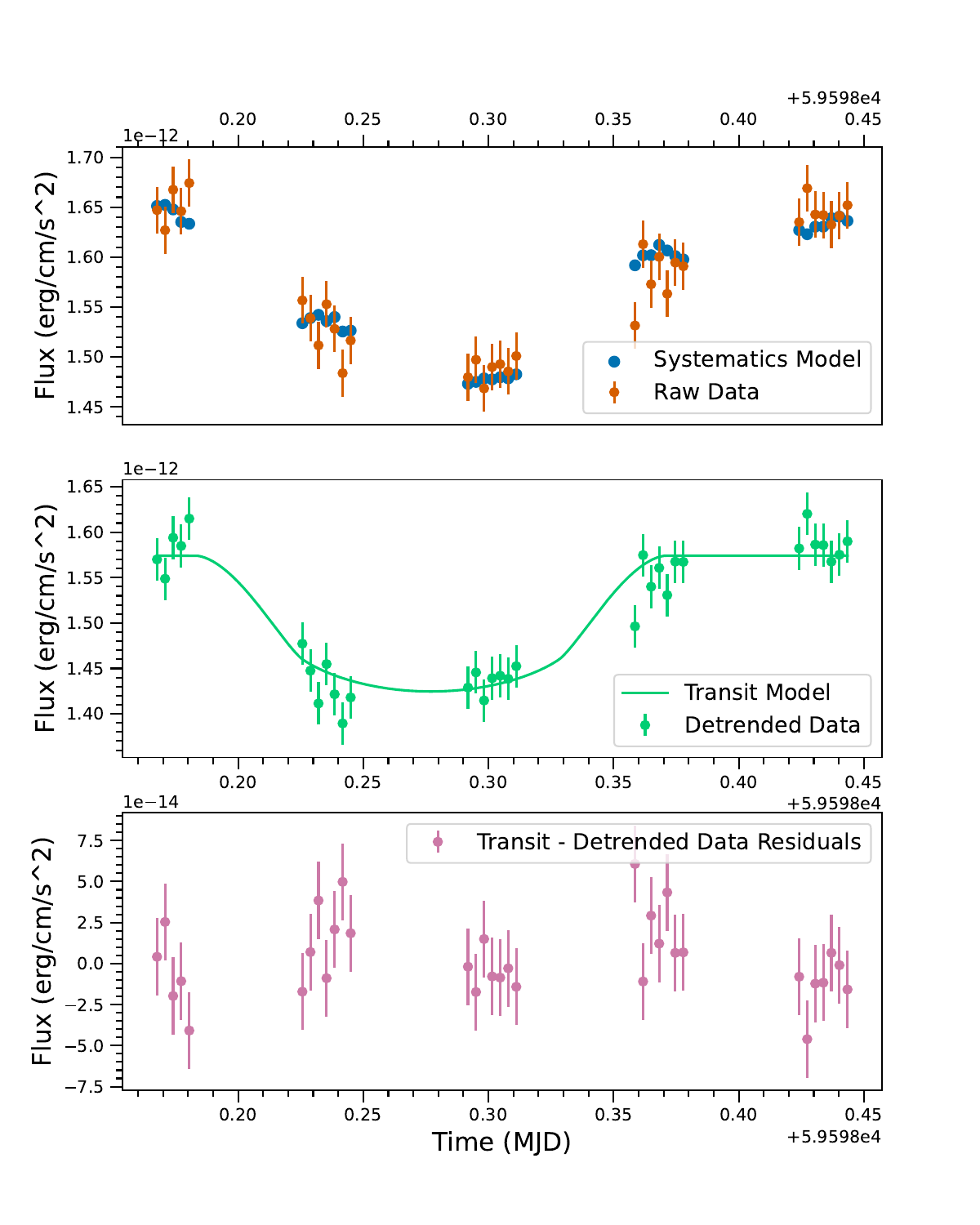}{0.52\textwidth}{Inside Mg II (2802.4 $\mathrm{\AA}$)}}
\caption{Same as Figure~\ref{kelt9b_spec_LC_v1}, but for two 1-$\mathrm{\AA}$ wide bins from Visit 2, respectively outside and inside the Mg II planetary absorption.}
\label{kelt9b_spec_LC_v2}
\end{figure*}

After fitting parameters for the white light curve, we created 1300 bins of wavelengths. Each bin covers 17 data points along each order of the echelle spectrum, corresponding to roughly 0.2~\rm{\AA}. We discarded 20 data points from either edge of each order due to low signal-to-noise.
Fortunately, the 2707\,$\rm \AA$ central wavelength setting for the E230M grating is designed such that the Mg II doublet and many of the Fe II line are comfortably away from any such overlap regions of sparse or noisy data. We then repeated the \texttt{curve\_fit} fitting process, fitting the apparent planetary radius $r_p$ and systematics parameters $a_n$ for each wavelength bin, while holding the transit center time $t_0$ steady. We show the fitted light-curve of bins inside and outside the Mg II doublet for Visits 1 and 2 in Figures~\ref{kelt9b_spec_LC_v1} and \ref{kelt9b_spec_LC_v2}, respectively.
\\

\begin{deluxetable}{lr|r}
    \tablecaption{Best-fit Parameters for Visits 1 and 2 \label{table:transit_best_fits}}
    \tablehead{\colhead{Parameter} & \colhead{Visit 1} & \colhead{Visit 2}}
    \startdata
    $t_0$ (MJD)  & $59479.7691 \pm 0.0026$   &  $59598.25955 \pm 0.0022$ \\
    ${R_p}/{R_s}$ & $0.11495 \pm 0.0087$  &  $0.08985\pm 0.0010$\\
    \enddata
\end{deluxetable}

\subsection{KELT-9b's Transmission Spectrum}

\begin{figure*}[ht!]
\plotone{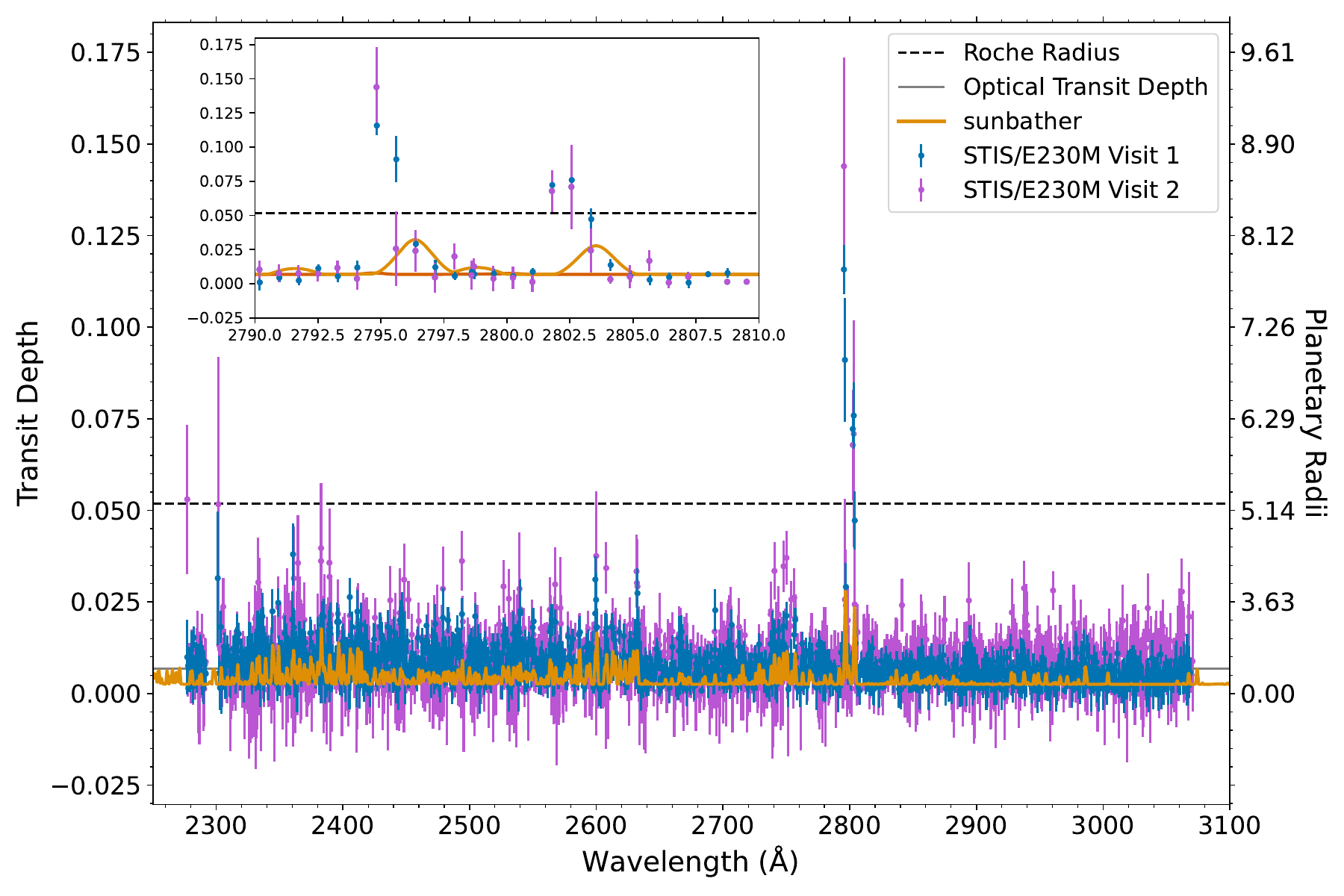}
\caption{Best-fit \texttt{sunbather} model (gold, $T=$13,200~K, $\dot{M}=10^{11.5}$ g/s, $V_{\mathrm{broad}} = 50$ km/s) to the combined two-visit dataset. The best-fit model to the entire spectrum does not match the magnitude or blueshift of the Mg II absorption near 2,800~$\mathrm{\AA}$, shown in the inset plot, which extends beyond the Roche radius (black, dashed).
\label{fig:modelfit}}
\end{figure*}

The resulting transmission spectrum of KELT-9b is shown in Figure~\ref{fig:modelfit}. We find clear evidence of increased absorption during the planet's transit by the atmosphere of KELT-9b, in particular the Mg II doublet at 2797 and 2803 $\mathrm{\AA}$, and the Fe II lines around 2600 $\mathrm{\AA}$. The depth of the Mg II lines at $R_{\rm{p}}^2/R_{\rm{s}}^2\approx 0.1$ implies a radial extend that exceeds the planet's Roche radius at about $R_{\rm{p}}^2/R_{\rm{s}}^2\approx 0.05$ ($3.7\times10^{10}$ cm, 2.77 $R_{\rm{p}}$, or 5.23 $R_{\rm{J}}$), implying the escaping gas is no longer bound to the planet. The Mg II line depth also exceeds the predictions from hydrostatic NLTE atmosphere modeling \citep{fossati:2021} of $R_{\rm{p}}^2/R_{\rm{s}}^2\approx 0.0361$, again implying atmospheric escape and the need for a non-hydrostatic density profile.

\section{Atmosphere Modeling}\label{sec:methods:models}

Next, we use 1D NLTE atmospheric escape models of KELT-9b to fit these data to quantify the significance of the Mg II and Fe II detections, as well as constrain the mass-loss rate and escaping-gas kinematics. We used the \texttt{sunbather} code \citep{linssen:2024} to generate synthetic spectra to compare to the observations. \texttt{sunbather} first calculates an isothermal Parker wind model \citep{parker:1958,oklopcic:2018} generated from the \texttt{p-winds} code \citep{dosSantos:2022} for a given temperature and mass-loss rate. The Parker wind model assumes the outflow is completely ionized (for the purposes of the flow velocities - ion densities are re-calculated self-consistently later) and a uniform sound speed as a function of radial distance. MHD effects are not directly accounted for in this simplified model. The model is calculated with 1000 layers from 1 to 20 planetary radii and assumes a default hydrogen number fraction of 0.9. We refer to the temperature at which the isothermal \texttt{p-winds} model is calculated as the ``input temperature".

The densities and velocities of the Parker wind model are then input into the non-local thermodynamic equilibrium (NLTE) photoionization model \texttt{Cloudy} \citep{ferland:1998,chatzikos:2023,gunasekera:2023} via \texttt{sunbather}. \texttt{Cloudy} calculates the NLTE energy level populations of atoms up to Zn, including photoionization, and the electron density. We calculate models that assume the isothermal \texttt{p-winds} input temperature, as well as models with an temperature profile in radiative equilibrium. The radiative equilibrium models are iterated until the altitude-dependent temperature profile is in a steady state, balancing heating and cooling from radiation, adiabatic expansion (cooling of material through expansion at lower pressures), and advection (heat flow). We refer the reader to \cite{linssen:2024} for details on the implementation of these terms. The NLTE model is calculated out to eight planetary radii with a dynamic number of pressure levels ($\gtrsim100$) and assumes a solar metallicity gas composition. We used a NLTE $T_{\rm{eff}}=10000$~K PHOENIX photosphere model at the location of KELT-9b's orbit as our irradiation spectrum \citep{hauschildt:1999}. 

We ran \texttt{sunbather} with varying mass-loss rates from $\dot{M}=10^{10.0}$ to $10^{14.0}$ g/s in steps of 0.5 dex, assuming a 13,200~K input temperature, consistent with the measurement from \cite{wyttenbach:2020} of 13,200$^{+800}_{-700}$~K. Exploring temperature space was challenging due to trouble consistently converging the temperature profile in \texttt{Cloudy}, especially for input temperatures near or below 10,000~K. Our investigation showed that the non-convergence was due to high densities in the lower atmosphere. 

\begin{figure}
    \centering
    \includegraphics[width=0.9\linewidth]{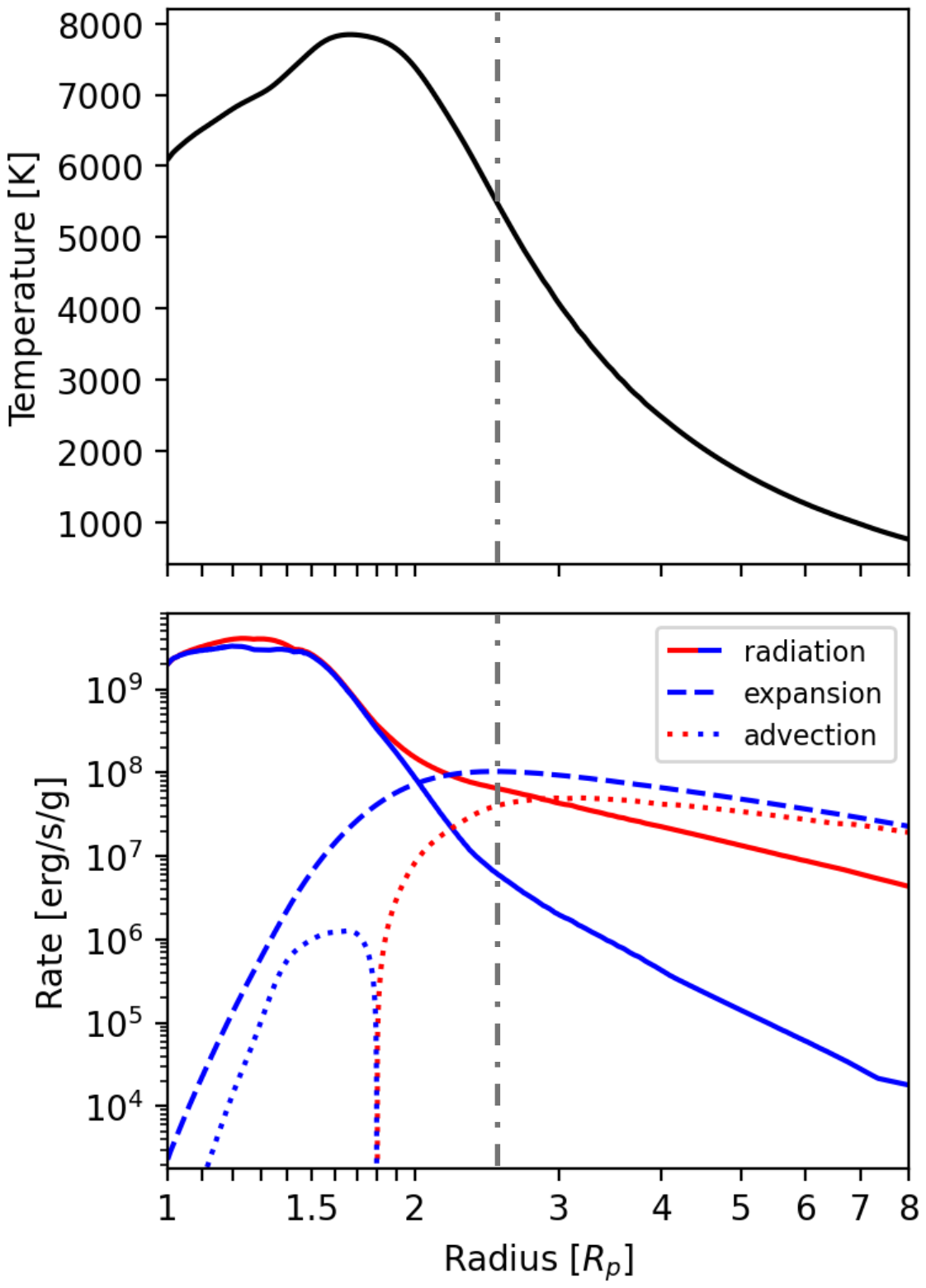}
    \caption{Top: Converged temperature profile of the best-fit Cloudy model to the Mg II feature, which had a input temperature of 13200~K and a mass-loss rate of $10^{12.5}$ g/s. Bottom: The heating (red) and cooling rates (blue) of the above model from radiation, expansion, and advection. The grey dash-dot line represents the location of the Roche radius.}
    \label{fig:cloudy_TP}
\end{figure}

We also found that the temperatures converged in \texttt{Cloudy} were much lower than the input 13,200~K. Figure~\ref{fig:cloudy_TP} shows the profile temperature profile, as well as the heating and cooling terms, for our best-fit model, described below. The relative importance of the various radiative heating and cooling terms qualitatively match the hydrostatic results of \cite{fossati:2021}, with metal line absorption, hydrogen photoionizations, and H- absorption important for heating and Mg and H lines important for cooling. The largest difference between our escaping models and the hydrostatic models is the importance of free-free interactions in cooling the atmosphere, perhaps due to the increased density from the outflowing gas. A deeper investigation and comparison would be helpful in evaluating the importance of different processes in setting the properties of the escaping atmosphere.

A maximum temperature of about 8,000~K is reached in our converged models, consistent with previous NLTE modeling \citep{fossati:2021}. As such, we ran an additional set of models with a fixed, isothermal temperature profile equal to the temperature of the \texttt{p-winds} input temperature. The lack of a more extreme model exospheric temperature for KELT-9b is in part due to the lack of high-energy radiation coming from what we assume is a chromospherically inactive early-type host star and cooling through expansion of the atmosphere through escape (see Figure~\ref{fig:cloudy_TP}).

We note a few more similarities and differences between our modeling approach and that of \cite{fossati:2021}. Both models consider a wide-range of all relevant atoms and ions and do not include molecular lines besides hydrogen molecules (e.g., H$_2$, H$_2^+$, H$_3^+$). Additionally, both models use a very similar input irradiation spectrum, both based on a 10,000~K PHOENIX stellar photosphere model. Our models extend out to 8 planetary radii to trace absorption well past the Roche lobe. On the other hand, the \cite{fossati:2021} model pressures below $10^{-11}$ bar, or roughly $<2-3 R_p$, where their assumption of a hydrostatic atmosphere are likely valid (since \cite{fossati:2021} are primarily investigating optical and infrared lines). Below $10^{-4}$ bar, \cite{fossati:2021} use a modified version of the 1D atmosphere model HELIOS \citep{malik:2019} to set a LTE lower-boundary condition for the NLTE upper atmosphere model. This significant difference in model domain and assumed structure limit inter-comparison with our own approach, though the fact that both models roughly agree on portions of the temperature structure and certain heating and cooling drivers shows the robustness of those results.

With the expected temperature, density, and level populations in hand, we then generated spectra that include atoms and ions (up to triple ionization) for all species included in \texttt{sunbather} using the NIST Atomic Spectra Databse\footnote{\url{https://www.nist.gov/pml/atomic-spectra-database}}: H, He, Li, Be, B, C, N, O, F, Ne, Na, Mg, Al, Si, P, S, Cl, Ar, K, Ca, Ca, Sc, Ti, V, Cr, Mn, Fe, Co, Ni, Cu, and Zn. We also generate spectra each with and without Fe II and Mg II lines, which are expected to be the main absorbers at the relevant wavelengths, temperatures, and pressures.

By eye, the observed line widths of the Mg II lines are much wider than expectations from thermal broadening alone as discussed in Section~\ref{sec:results:turb}, so we extended the grid of spectrum calculations to include a broadening velocity ($V_{\mathrm{broad}}$) at 0, 25, 50, 75, and 100 km/s, included in the \texttt{sunbather} models as an additional term in the standard deviation of the Gaussian component to the Voigt line profile of all lines. This serves to broaden the line, while also diluting the line core. In the end, we have two small grids, one with an isothermal temperature at 8000~K and another with a converged temperature profile with an input temperature of 13,200~K. We refer to these grids as the $T_\mathrm{iso}$=8,000~K grid and the $T_\mathrm{input}$=13,200~K grid, respectively. Each grid has two free parameters, $\dot{M}$ and $V_{\mathrm{broad}}$.

We then fit the individual models to the observations (not using any interpolation between model grid-points) using a least-squares regression with an additional offset parameters in the mean transit depth and the Doppler shift in velocity ($\lambda_{\rm shift}$). The models are first convolved to the instrumental resolution and then the flux across each pixel is averaged. For the model including all species, we reach $\chi^2/N$, of 1.45 ($N = 1228$) and 1.36 ($N = 1199$) for the first and second visit, respectively, where $N$ is the number of data points. When fitting each visit jointly with the same model, we reach a worse $\chi^2/N$ of 2.07, implying different model parameters are preferred for each visit. Fits with the models that do not include Mg II or Fe II fit uniformly worse for each visit, including the combined dataset. The best-fit values for every scenario are shown in Table~\ref{table:mass_loss_turbulence_chi} and we evaluate these fits in Section~\ref{sec:results}.

To further investigate the kinematics of KELT-9b's atmosphere, we also performed direct Gaussian line fitting to the spectrum, with particular attention paid to the Mg II doublet. We fit for the Doppler shift, the $\sigma$ line-width, and the maximum depth of the line. This was also fit with a least-squares regression, with the resulting quantities listed in Table~\ref{tab:spectral}. We evaluate these fits in Section~\ref{sec:results}.

\section{Results} \label{sec:results}

\begin{deluxetable*}{lccccc}
\tablecaption{Best-fit Mass-Loss Rate, Velocity Broadening, and $\chi^2$/N for our two model grids: $T_\mathrm{input}$=13,200~K, and $T_\mathrm{iso}$=8,000~K \label{table:mass_loss_turbulence_chi}}
\tablehead{
    \colhead{Parameter} & \colhead{All Species} & \colhead{Only Mg II} & \colhead{No Mg II} & \colhead{Only Fe II} & \colhead{No Fe II} 
}
\startdata
\multicolumn{6}{c}{\textbf{$T_\mathrm{iso}$=8,000~K Grid}} \\
\hline
\multicolumn{6}{c}{\textbf{Visit 1} (N=1228)} \\
$\log_{10} \dot{M}$ (g/s) & 12.0 & 10.5 & 11.5 & 10.5 & 10.5\\
$V_{\mathrm{broad}}$(km/s) & 75.0 & 75.0 & 50.0 & 50.0 & 50.0\\
$\lambda_{\mathrm{shift}}$(\AA) & $0.22 \pm 0.04$ & $-9.1 \pm 50.6$ & $0.42 \pm 0.034$ & $0.24 \pm 0.05$ & $0.72 \pm 0.072$\\
$\chi^2$/N & 1.513 & 1.169 & 1.661 & 1.722 & 2.060\\
$\Delta\chi^2$\tablenotemark{$^\dagger$} & -- & $-422.2$ & $182.2$ & $256.7$ & $672.3$\\
\hline
\multicolumn{6}{c}{\textbf{Visit 2} (N=1199)} \\
$\log_{10} \dot{M}$ (g/s) & 8.0 & 13.5 & 8.0 & 10.5 & 9.0\\
$V_{\mathrm{broad}}$(km/s) & 50.0 & 25.0 & 50.0 & 0.0 & 25.0\\
$\lambda_{\mathrm{shift}}$(\AA) & $0.51 \pm 0.06$ & $-0.99 \pm 0.39$ & $0.51 \pm 0.07$ & $0.497 \pm 0.008$ & $0.24 \pm 0.05$\\
$\chi^2$/N & 1.428 & 0.356 & 1.437 & 1.402 & 1.499\\
$\Delta\chi^2$\tablenotemark{$^\dagger$} & -- & $-1285.8$ & $10.3$ & $-31.2$ & $85.2$\\
\hline
\multicolumn{6}{c}{\textbf{Combined} (N=2427)} \\
$\log_{10} \dot{M}$ (g/s) & 9.0 & 13.0 & 9.0 & 11.0 & 9.5\\
$V_{\mathrm{broad}}$(km/s) & 25.0 & 50.0 & 25.0 & 0.0 & 25.0\\
$\lambda_{\mathrm{shift}}$(\AA) & $0.45 \pm 0.02$ & $-0.35 \pm 0.035$ & $0.46 \pm 0.02$ & $0.508 \pm 0.005$ & $0.30 \pm 0.03$\\
$\chi^2$/N & 2.322 & 2.207 & 2.345 & 2.303 & 2.527\\
$\Delta\chi^2$\tablenotemark{$^\dagger$} & -- & $-277.5$ & $57.6$ & $-46.0$ & $498.2$\\
\hline
\multicolumn{6}{c}{\textbf{$T_\mathrm{input}$=13,200~K Grid}} \\
\hline
\multicolumn{6}{c}{\textbf{Visit 1} (N=1228)} \\
$\log_{10} \dot{M}$ (g/s) & 12.0 & 13.0 & 12.0 & 12.0 & 12.5\\
$V_{\mathrm{broad}}$(km/s) & 75.0 & 100.0 & 50.0 & 50.0 & 75.0\\
$\lambda_{\mathrm{shift}}$(\AA) & $0.146 \pm 0.042$ & $0.5 \pm 0.5$ & $0.290 \pm 0.033$ & $0.237 \pm 0.034$ & $-0.11 \pm 0.05$\\
$\chi^2$/N & 1.464 & 1.172 & 1.666 & 1.688 & 2.003\\
$\Delta\chi^2$\tablenotemark{$^\dagger$} & -- & $-358.3$ & $248.3$ & $275.2$ & $661.8$\\
\hline
\multicolumn{6}{c}{\textbf{Visit 2} (N=1199)} \\
$\log_{10} \dot{M}$ (g/s) & 11.5 & 10.0 & 11.5 & 11.5 & 12.0\\
$V_{\mathrm{broad}}$(km/s) & 25.0 & 100.0 & 25.0 & 25.0 & 75.0\\
$\lambda_{\mathrm{shift}}$(\AA) & $0.322 \pm 0.012$ & $4.35 \pm 4.35$ & $0.322 \pm 0.013$ & $0.328 \pm 0.013$ & $-0.27 \pm 0.08$\\
$\chi^2$/N & 1.369 & 0.375 & 1.383 & 1.399 & 1.477\\
$\Delta\chi^2$\tablenotemark{$^\dagger$} & -- & $-1191.3$ & $17.2$ & $35.4$ & $129.6$\\
\hline
\multicolumn{6}{c}{\textbf{Combined} (N=2427)} \\
$\log_{10} \dot{M}$ (g/s) & 11.5 & 12.5 & 11.5 & 11.5 & 12.5\\
$V_{\mathrm{broad}}$(km/s) & 50.0 & 50.0 & 50.0 & 50.0 & 75.0\\
$\lambda_{\mathrm{shift}}$(\AA) & $0.454 \pm 0.029$ & $-0.531 \pm 0.046$ & $0.51 \pm 0.03$ & $0.51 \pm 0.04$ & $-0.30 \pm 0.04$\\
$\chi^2$/N & 2.117 & 1.927 & 2.174 & 2.206 & 2.334\\
$\Delta\chi^2$\tablenotemark{$^\dagger$} & -- & $-461.0$ & $137.6$ & $214.5$ & $525.8$\\
\enddata
\tablenotetext{\dagger}{Because the number of free parameters is the same in these comparisons, $\Delta\chi^2 = \Delta BIC$.}
\end{deluxetable*}

\subsection{Detection of Mg II and Fe II}

To quantify the detection significance of Mg II and Fe II, we first compare our results with the fiducial grid of spectra (which included all species) to grids without Fe II and Mg II, individually. This comparison allowed us to quantify the detection significance of each ion. This was done by simply removing the relevant opacity source during the spectral synthesis step. Because metals and their ions have an important effect on the heating and cooling rates \citep{fossati:2021}, and therefore the temperature structure, we only consider the tests on the 8000~K isothermal grid, where temperature is controlled for.

Table~\ref{table:mass_loss_turbulence_chi} shows the resulting best fit model parameters and their $\chi^2$ values from fits across the entire wavelength range. Because the number of free parameters is not changing  between the models, only which species are included in the opacity calculation, the change in $\chi^2$ corresponds to the $\Delta BIC$, or the Bayesian Information Criterion, where anything larger than 10 is considered a highly significant detection \citep{raftery:1995,liddle:2007}. 
Fits to the first, second, and combined-visit data each indicate a highly significant detection of Mg II with $\Delta BIC > 10$ in all cases.

A similar story holds for Fe II, though Fe II absorbs in two bands at 2400 and 2600 $\mathrm{\AA}$ rather than a more obvious doublet like Mg II. Because Fe II absorbs over a much wider wavelength range than Mg II, it affects the model fit to a higher degree and thus each of the visits detects Fe II at an even higher significance than Mg II, despite Mg II being more identifiable by eye.

\begin{figure}[ht!]
\includegraphics[width=0.48\textwidth]{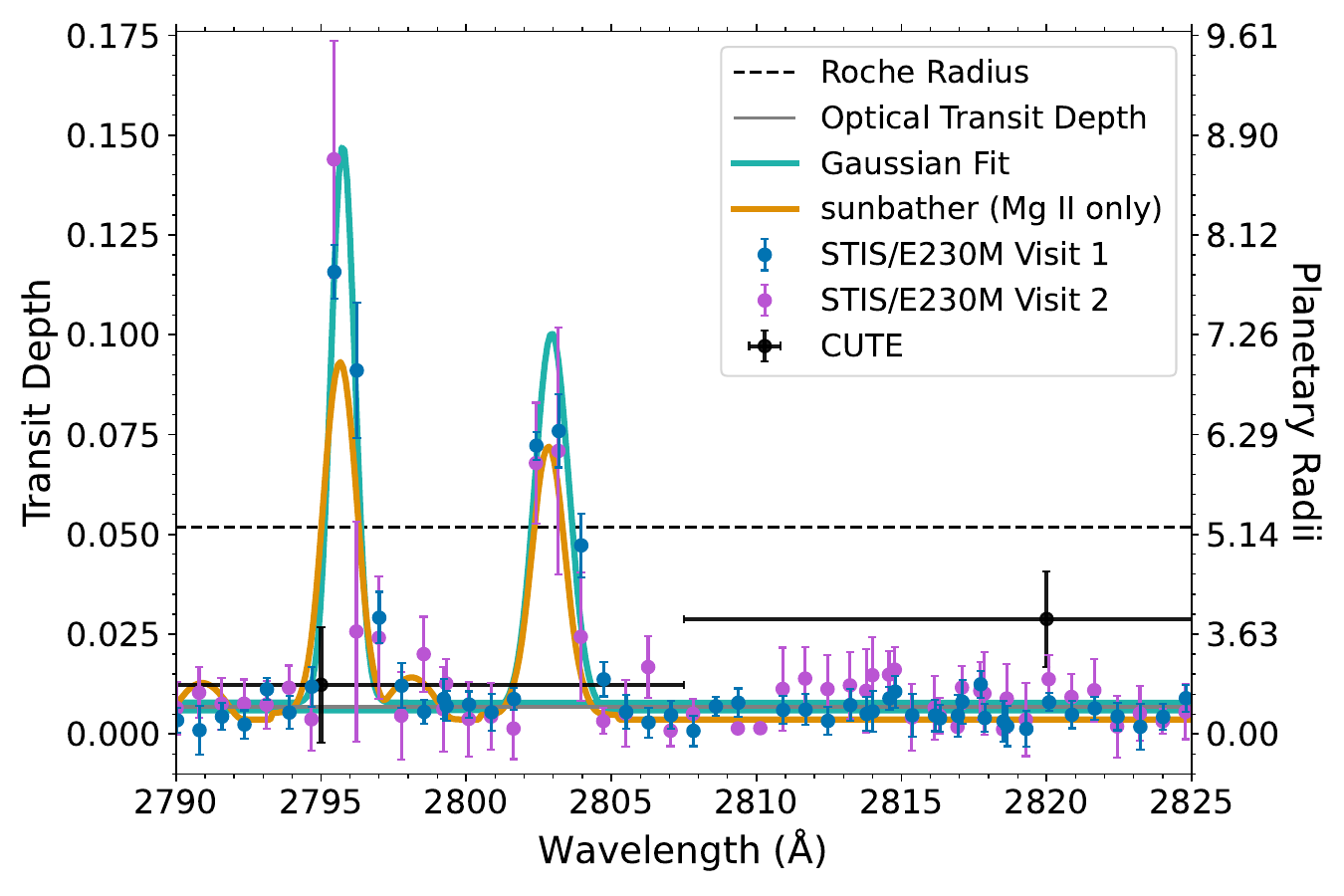}
\caption{Best-fit model spectrum (gold, $T=$13,200~K, $\dot{M}=10^{12.5}$ g/s, $V_{\mathrm{broad}} = 50$ km/s) to the 100 data points closest to the Mg II doublet from visit 1 (blue) and 2 (purple). A Gaussian fit to the Mg II is shown in green. Also plotted are the observations of KELT-9b from the Colorado Ultraviolet Transit Experiment \citep[CUTE, black and grey, visit 1 and 2, respectively,][]{egan:2024}. The HST/STIS observations are consistent with the CUTE observations when binned to the same wavelength scale.
\label{fig:modelfit:MgII}}
\end{figure}

\subsection{Mass-loss Rate}

Table~\ref{table:mass_loss_turbulence_chi} also shows the best-fit values for mass-loss rate ($\dot{M}$) and the line-broadening velocity ($V_{\mathrm{broad}}$). In general, we see high mass-loss rates from the $T_\mathrm{input}$=13,200~K grid compared to the $T_\mathrm{iso}$=8,000~K grid. The fits from the $T_\mathrm{iso}$=8,000~K grid to the Visit 2 spectrum are much smaller than both Visit 1 or the $T_\mathrm{input}$=13,200~K grid. The $T_\mathrm{iso}$=8,000~K grid only finds a high mass loss for Visit 2 and the combined data set when Mg II is the only species considered.

\begin{figure*}
\centering
\gridline{\fig{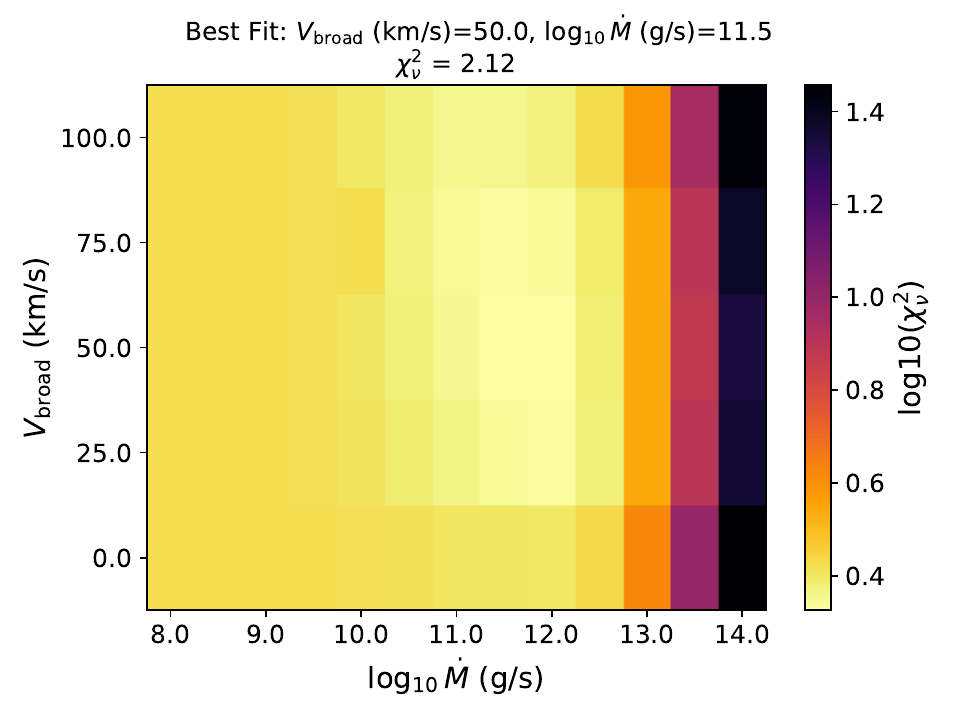}{0.5\textwidth}{(a) All species, $T_\mathrm{input}$=13,200~K}\fig{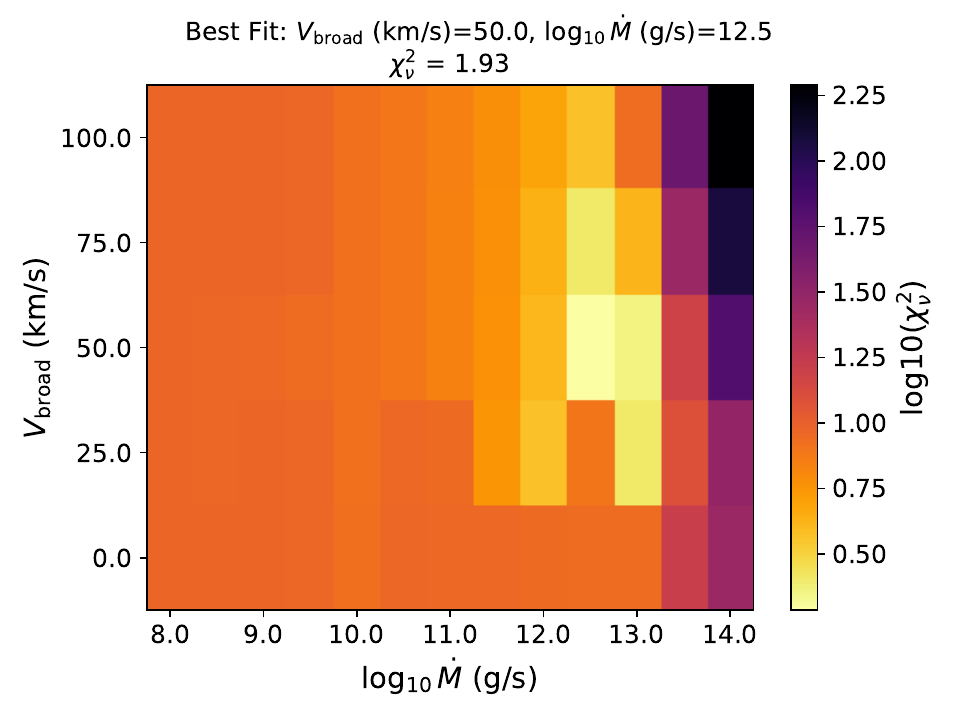}{0.5\textwidth}{(b) Mg II only, $T_\mathrm{input}$=13,200~K}}
\gridline{\fig{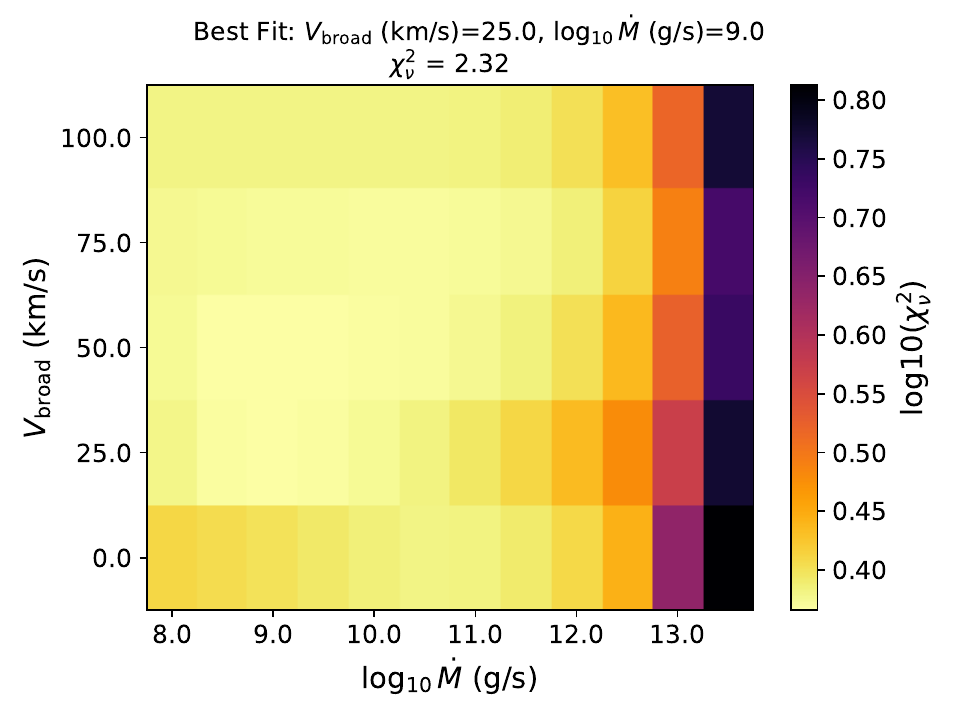}{0.5\textwidth}{(c) All species, $T_\mathrm{iso}$=8,000~K}\fig{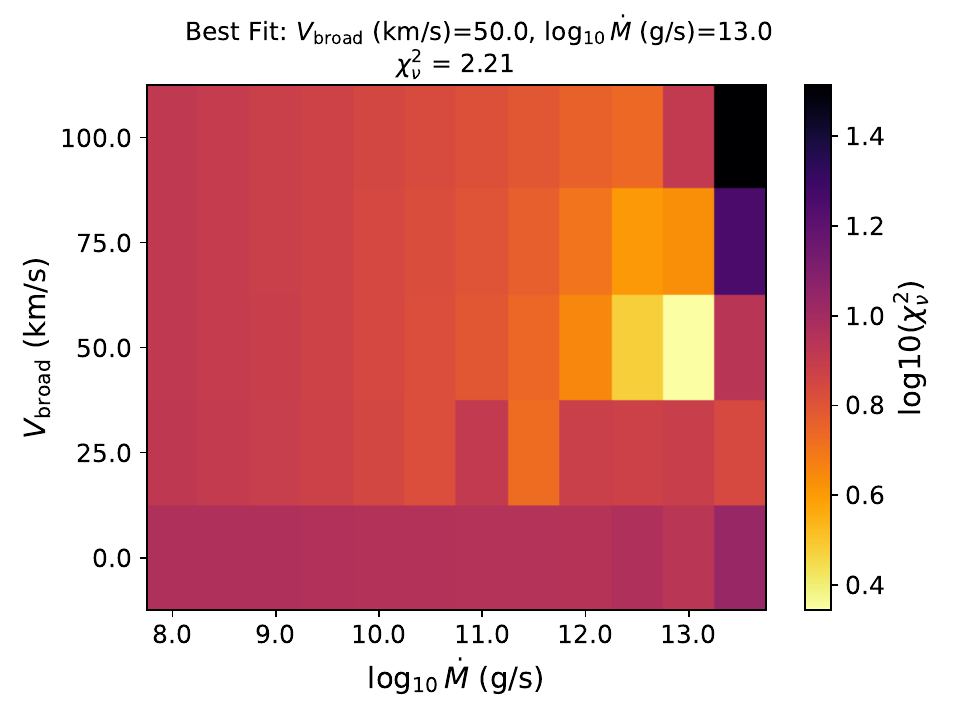}{0.5\textwidth}{(d) Mg II only, $T_\mathrm{iso}$=8,000~K}}
\caption{$\chi^2/N$ grids of the mass-loss rate ($\log_{10} \dot{M}$) versus the broadening velocity ($V_\mathrm{broad}$) for the two-visit combined data set. The left plot shows the models with all species included fit to the whole wavelength range. The right plot shows the Mg II-only model fit to the data between 2783 and 2818 $\mathrm{\AA}$.\label{fig:chigrid}}
\end{figure*}

Figure~\ref{fig:chigrid} shows how the $\chi^2/N$ of the combined dataset varies as $\dot{M}$ and $V_{\mathrm{broad}}$ change for the $T_\mathrm{input}$=13,200~K and the $T_\mathrm{iso}$=8,000~K grids. $\dot{M} = 10^{11.5}$~g/s was preferred for the $T_\mathrm{input}$=13,200~K grid, which was much higher than the $\dot{M} = 10^{9.0}$~g/s found for the $T_\mathrm{iso}$=8,000~K grids grid. A clear upper limit is apparent at $10^{13.0}$ g/s for both grids, where the fits worsen dramatically.

When just considering the 100 data points nearest the Mg II feature, much higher $\dot{M}$ values are found with $10^{12.5}$ and $10^{13.0}$ g/s found for the $T_\mathrm{input}$=13,200~K and the $T_\mathrm{iso}$=8,000~K grids, respectively. The best-fit Mg II mass-loss rates are very close to the mass-loss rate measured from the hydrogen Balmer lines in \cite{wyttenbach:2020}, which found $\dot{M} = 10^{12.8\pm 0.3}$~g/s, and is within the $10^{12} - 10^{13}$ g/s range found in \cite{shaikhislamov:2025}. Fits to the O I triplet at 7774~$\mathrm{\AA}$ were slightly lower at $10^{11} - 10^{12}$ g/s. We discuss this further in Section~\ref{sec:disc:previous}.

\subsection{Line Broadening}\label{sec:results:turb}

As stated in Section~\ref{sec:methods:models}, thermal broadening alone ($\sim 5.0$ km/s at 13,200~K) is insufficient to explain the observed widths of Mg II and Fe II lines. The models used to fit the Mg II and Fe II lines here require a velocity broadening, $V_{\mathrm{broad}}$, added as an additional Gaussian component to the lines' Voigt profile. 
$V_{\mathrm{broad}}\approx25-75$ km/s is preferred to fit the data. The only scenarios where no additional broadening is needed is when only Fe II opacity is considered in the $T_\mathrm{iso}$=8,000~K grid. 

This high velocity is also suggested by Gaussian fits to the Mg doublet (see Table~\ref{tab:spectral}), with a FWHM of 150.11 $\pm$ 8.07 and 180.40 $\pm$ 8.44 km/s (1.32 $\pm$ 0.08 and 1.61 $\pm$ 0.08 $\mathrm{\AA}$) measured for Mg II H and K, respectively, in the combined two-visit dataset. This corresponds to a velocity of 63.75 $\pm$ 8.07 and 76.61 $\pm$ 8.44 km/s, respectively. 

This level of $V_{\mathrm{broad}}$ is unexpectedly large and the mechanism driving this broadening is not clear. Previous observations of individual atomic lines also required extra line broadening, though at much lower magnitudes, around 10-20 km/s \citep{borsa:2022,DArpa24}. We show some relevant velocities for the KELT-9 system in Table~\ref{tab:velocities}. For example, the thermal broadening of Mg at 13,200~K is about 5.0 km/s, which is far too low to explain the line widths we observe. KELT-9b's rotational velocity at its optical transit radius is approximately 6.6 km/s, assuming tidal locking. For two hemispheres moving in opposite directions, that accounts for 13.2 km/s of broadening. For material co-rotating at the Roche radius of 5.23 R$_J$, the velocity would be about 18.3 km/s from each hemisphere, or 36.6 km/s total, similar to the value for the Doppler shift, but less than the necessary broadening velocity. Meanwhile, for reference, KELT-9b's orbital velocity is about 257 km/s, and the stellar rotation is measured to be $v\sin{i} = 114.9 \pm 3.4$ km/s \citep{kama:2023}.

Photochemical models with dynamics suggest outflows can reach up to 30 km/s for the highest irradiation fluxes tested (100$\times$ solar average flux) in \cite{koskinen:2012a}. A stellar wind may also accelerate escaping gas; the solar wind at Earth's orbital location varies between about 200 and 600 km/s over the course of the year\footnote{https://www.swpc.noaa.gov/products/real-time-solar-wind}, which could be more than sufficient to accelerate escaping ions if KELT-9b experiences a similar stellar wind. However, KELT-9 may be hot enough to not exhibit chromospheric activity \citep{fossati:2018} generally thought to drive traditional solar-like winds. Radiation pressure could also accelerate metal atoms to $\approx100 km/s$ very quickly \cite{cauley:2019}. Multi-dimensional modeling of this and other escaping lines of KELT-9b can likely provide a more complete picture of the physical process at work in the KELT-9 system.

\subsection{Doppler Shift}

We also measure a significant blueshift in the NUV spectrum. From the best-fit $T_\mathrm{input}$=13,200~K Mg II-only atmosphere model, we find a net blueshift of $-0.531 \pm 0.05$ $\mathrm{\AA}$, corresponding to $57 \pm 5$~km/s. For the corresponding $T_\mathrm{iso}$=8,000~K model, we find a net blueshift of $-0.350 \pm 0.035$ $\mathrm{\AA}$, corresponding to $37 \pm 4$~km/s. The fits from models including other species (chiefly Fe II) do not find a consistent blue shift, but vary between a redshift of 0.1 to about 0.5 $\mathrm{\AA}$. Here, we focus on the clearly resolved Mg II lines.

The Gaussian fits find a net blueshift of $-0.35 \pm 0.01$ and $-0.36 \pm 0.01$ $\mathrm{\AA}$ for Mg II H \& K, respectively, in the combined two-visit dataset. This corresponds to $-37.6 \pm 1.0$ and $-38.0 \pm 0.7$~km/s, respectively. This is close to the blueshift values found from the atmosphere models, especially the $T_\mathrm{iso}$=8,000~K grid. These results suggest that the escaping Mg II is moving away from the star at a net velocity exceeding the sonic speed of about 13.4 km/s at $T=13200$~K. The escaping gas may be blown away by the stellar wind \citep{mccann:2019,carolan:2021}, though the early-A type host star may exhibit a radiation pressure driven wind that may be responsible for the acceleration \cite{cauley:2019}. 

To verify the quality of HST's wavelength solution and the reliability of the system radial velocity, we measured the line position of Mg II in the stellar spectrum shown in Figure~\ref{fig:stellar_spec}. For Mg II H \& K, we find positions of 2796.25 and 2803.39~$\mathrm{\AA}$, respectively, corresponding to blueshifted velocities of $-15.9$ and $-19.9$~km/s, which both agree with the measured system velocity of $-17.7$~km/s \citep{wyttenbach:2020} to within the accuracy of STIS's wavelength calibration at $\pm5$~km/s .\footnote{The quoted absolute wavelength accuracy of STIS is $\approx 0.5-1$~pixel\footnote{https://hst-docs.stsci.edu/stisihb/chapter-16-accuracies/16-1-summary-of-accuracies}. With the E230M dispersion of $\sim60,000$ $\mathrm{\AA}$ per pixel, the absolute wavelength precision should be better than 0.05~$\mathrm{\AA}$, or 5 km/s, at 2800~$\mathrm{\AA}$.}

\begin{deluxetable*}{lccccc}
\tablecaption{Gaussian Fit Measurements to the Mg II Doublet Compared to Select Previous Line Measurements \label{tab:spectral}}
\tablehead{
    \colhead{Visit} & \colhead{Doppler Shift (\AA)\tablenotemark{a}} & \colhead{Doppler Shift (km/s)\tablenotemark{a}} & \colhead{FWHM (\AA)} & \colhead{FWHM (km/s)} & \colhead{$V_{\mathrm{broad}}$ (km/s)}
}
\startdata
\multicolumn{6}{c}{\textbf{Mg II H} ($\lambda_{\mathrm{vac}} = 2796.35$ \AA)} \\
\hline
Visit 1 & $-0.31 \pm 0.01$ & $-32.73 \pm 1.22$ & $1.39 \pm 0.08$ & $157.41 \pm 8.79$ & $66.85 \pm 8.79$ \\
Visit 2 & $-0.29 \pm 0.01$ & $-31.51 \pm 0.99$ & -- & -- & -- \\
Combined & $-0.35 \pm 0.01$ & $-37.58 \pm 1.08$ & $1.32 \pm 0.08$ & $150.11 \pm 8.07$ & $63.75 \pm 8.07$ \\
\hline
\multicolumn{6}{c}{\textbf{Mg II K} ($\lambda_{\mathrm{vac}} = 2803.53$ \AA)} \\
\hline
Visit 1 & $-0.32 \pm 0.01$ & $-33.72 \pm 0.68$ & $1.70 \pm 0.09$ & $191.48 \pm 9.79$ & $81.32 \pm 9.79$ \\
Visit 2 & $-0.40 \pm 0.02$ & $-42.61 \pm 2.65$ & $1.40 \pm 0.23$ & $174.28 \pm 24.68$ & $74.01 \pm 24.68$ \\
Combined & $-0.36 \pm 0.01$ & $-38.03 \pm 0.66$ & $1.61 \pm 0.08$ & $180.40 \pm 8.44$ & $76.61 \pm 8.44$ \\
\hline
\multicolumn{6}{c}{\textbf{H$\alpha$} ($\lambda_{\mathrm{vac}} = 6564.6$ \AA)} \\
\hline
\cite{yan:2018} & $0.022\pm0.022$ & $-1.02^{+0.99}_{-1.00}$ & $1.17 \pm 0.04$ & $55.38 \pm 1.93$ & $23.52 \pm 1.93$ \\
\cite{cauley:2019} & -- & $2.67 \pm 0.28$ & -- & $55.1 \pm 0.6$ & --\\
\hline
\multicolumn{6}{c}{\textbf{H$\alpha, \beta, \gamma, \delta, \epsilon$}}\\
\hline
\cite{wyttenbach:2020} & $0.005\pm0.013$ & $0.2\pm0.6$ & $0.985 \pm 0.031$ & $44.9 \pm 1.4$ & $19.1 \pm 1.4$ \\
\cite{DArpa24} & -- & $-4.7\pm0.28$ & -- & $38.86 \pm 9.70$ & $16.50 \pm 4.12$ \tablenotemark{b}\\
\hline
\multicolumn{6}{c}{\textbf{O I Triplet} ($\lambda_{\mathrm{vac}} \approx 7774$ \AA)}\\
\hline
\cite{borsa:2022} & -- & -- & $0.793 \pm 0.3$ & $30.55 \pm 11.75$ & $13.0 \pm 5.0$ \tablenotemark{b}\\
\hline
\multicolumn{6}{c}{\textbf{Mg I} ($\lambda_{\mathrm{vac}} \approx 5167$--$5183, 5528$ \AA)}\\
\hline
\cite{cauley:2019} & -- & $-0.1 \pm 0.8$ & -- & $32.47 \pm 3.51$ & --\\
\cite{DArpa24} & -- & -- & $0.80 \pm 0.35$ & $45.92 \pm 8.55$ & $19.50 \pm 3.63$ \tablenotemark{b}\\
\enddata
\tablenotetext{a}{Doppler shift after subtracting the system radial velocity of $-0.1657\mathrm{\AA}$ or $-17.74$ km/s \citep{wyttenbach:2020}.}
\tablenotetext{b}{We quote here the macroturbulent velocity.}
\tablenotetext{}{\cite{DArpa24} finds all lines blueshifted between 0-10 km/s.}
\end{deluxetable*}

\begin{deluxetable*}{lcl}
\tablecaption{Relevant Velocities in the KELT-9 System\label{tab:velocities}}
\tablehead{
    \colhead{Velocity} & \colhead{Value [km s$^{-1}$]} & \colhead{Reference}
}
\startdata
Mg thermal velocity at 13,200~K & 5.0 & \\
Sonic Speed at 13,200~K & 13.4 & \\
Planetary rotational velocity & 6.6 & \citet{gaudi:2017} \\
Planetary rotational broadening at Roche radius & 18.3 & \citet{gaudi:2017} \\
Planetary Escape Velocity & 73.5 & \citet{gaudi:2017} \\
Planetary orbital velocity  & 257 & \citet{gaudi:2017} \\
System Radial Velocity & $-17.74 \pm 0.04$ & \citet{wyttenbach:2020} \\
Stellar projected rotational velocity ($v_{\rm *}\sin{i}$) & $116.9 \pm 1.8$ & \citet{wyttenbach:2020} \\
Stellar rotational velocity ($v_{\rm eq}$) & $195^{+37}_{-23}$ & \citet{wyttenbach:2020} \\
Solar wind velocity & 300--600 & \citet{larrodera:2020} \\
A-star radiation pressure wind & $\sim$100 & \cite{cauley:2019} \\
\hline
Mg II H $V_\mathrm{broad}$ (combined data set) & $63.75 \pm 8.07$ & This Work \\
Mg II K $V_\mathrm{broad}$ (combined data set) & $76.61 \pm 8.44$& This Work \\
\enddata
\end{deluxetable*}

\section{Discussion}\label{sec:disc}

\subsection{Comparison to Previous KELT-9b Observations} \label{sec:disc:previous}

\begin{figure*}
\includegraphics[width=0.95\textwidth]{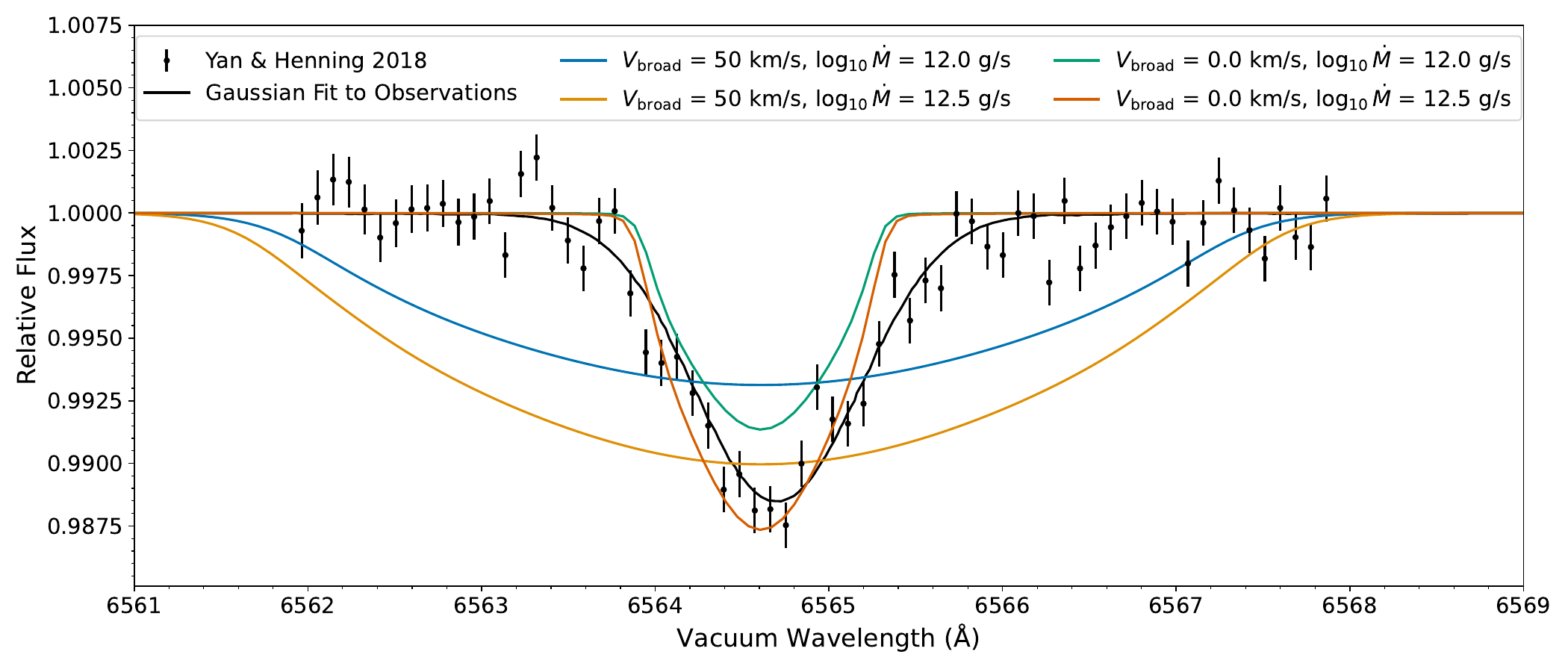}
\caption{Comparison between the H$\alpha$ observations of \cite{yan:2018}, their Gaussian fit to the line, and the extrapolation of our best-fitting models to optical wavelengths. While the best-fit mass-loss rate of 10$^{12.5}$ g/s to the Mg II line fits the depth of the H$\alpha$ line well, it requires a much narrower line with $V_{\mathrm{broad}}\approx 0$ km/s.
\label{fig:Halpha}}
\end{figure*}

This is not the first detection of significant atmospheric escape from KELT-9b. Indeed, since its discovery, KELT-9b has been the subject of many atmospheric escape studies due to its uniquely extreme temperature. As mentioned above, \cite{wyttenbach:2020} found escaping Balmer lines at a similar mass-loss rate as our observations indicate. It is interesting that H, Fe, and Mg atoms are escaping at similar rates, perhaps indicating a coupled outflow where heavy elements are entrained in the escaping gas \citep{koskinen:2012a,koskinen:2012b}. This is especially interesting since the Balmer lines may not be directly probing the hydrodynamic part of the atmosphere \citep{turner:2020} and that a hotter temperature profile from NLTE effects may also explain the line depths \citep{fossati:2021}.

If the Balmer H mass-loss esimates are accurate, it is also significant that \emph{ions} of Fe and Mg are escaping at a similar rate as neutral H. Any magnetic field on KELT-9b would be expected to significantly affect both the rate and morphology of escaping ions relative to neutral atoms, so the fact that we find similar escape rates for both neutral and ion species could suggest a weak magnetic field on KELT-9b, though the gas kinematics may tell a different story \citep{savel:2024}. As for the escape morphology, there is little direct evidence for a late egress, which would signal asymmetric escape, perhaps similar to the comet-like tail seen in the Neptune-size GJ 436b \citep{ehrenreich:2015}. \cite{Lowson:2023} recently characterized the shape of the H$\alpha$ absorption, finding a shape that deviated from the traditional transit shape towards more of a ``W" shape, which would be consistent with material escaping towards the observer. One might expect that a planetary magnetic field would change this morphology by confining ions to travel along field lines.

Unfortunately, the systematics and phasing of the HST orbits makes it difficult to characterize the light curve shape. Our five-orbit visits are also the longest possible with the MAMA detectors aboard HST due to the need to avoid the South Atlantic Anomaly (SAA), so attempts to measure a delayed egress would be difficult. Splitting observations up into multiple orbits interrupted by an SAA passage is also difficult due to the non-continuous systematic trends. However, our observations do indicate that the metal line absorption has at least ended by our fifth orbit, suggesting any cometary tail probed through Mg II and Fe II may be confined to be relatively near the planet.

One major difference between the absorption measured here and previously measured Balmer lines is our need for a large velocity broadening to fit the line profiles of Mg II. This indicates some process broadening the NUV metal ion lines much more significantly than the hydrogen Balmer lines. This could be simply due to the difference in pressure level probed, where the Balmer lines are seeing much deeper in the atmosphere compared to the Mg II doublet. The line widths we measure here are more similar to the Lyman-$\alpha$ line widths measured in, e.g., HD 209458b \citep{vidal-madjar:2003}, which probe gas accelerated by either radiation pressure or the stellar winds \citep{holmstroem:2008}. Self-consistent 3-dimensional modeling of the escape, building on the 1D modeling done here, may be able to elucidate the process behind this difference.

The similarities and differences with the Balmer absorption are borne out in Figure~\ref{fig:Halpha}, which shows an extrapolation of our best-fit models to the H$\alpha$ data from \cite{yan:2018}, both with and without the best-fit broadening velocity. With $V_\mathrm{broad} = 0$~km/s, the model fits surprisingly well, roughly matching the depth of the Gaussian best-fit to the data. However, the level of broadening is clearly different between the NUV observations and the hydrogen Balmer observations, with our best-fit $V_{\mathrm{broad}}=50$ km/s in clear disagreement. \cite{yan:2018} find the line-width of H$\alpha$ to be 21.7$\pm$1.1 km/s (FWHM = 51.2 km/s). More recent observations of these and other atomic lines shows a similar story \citep[see Table~\ref{tab:spectral};][]{cauley:2019,wyttenbach:2020,borsa:2022,DArpa24}.

KELT-9b has also been observed by the Colorado Ultraviolet Transit Experiment (CUTE) \citep{egan:2024}. The 27.5 cm$^2$ effective area CUTE CubeSat telescope observed two transits of KELT-9b with its 2479-3306~$\mathrm{\AA}$ spectrograph at a resolution of 3.9~$\mathrm{\AA}$ \citep{france:2023}. While the observations showed a broadband increase in transit depth across the NUV compared to the optical transit depth, strong systematics reduced the sensitivity to individual spectral lines. We plot the KELT-9b transit spectrum from CUTE in Figure~\ref{fig:modelfit:MgII} and find that if we bin the STIS observations to the 25~$\mathrm{\AA}$ bins of the presented CUTE spectrum, our observations are consistent. 

\begin{figure*}
\includegraphics[width=0.48\textwidth]{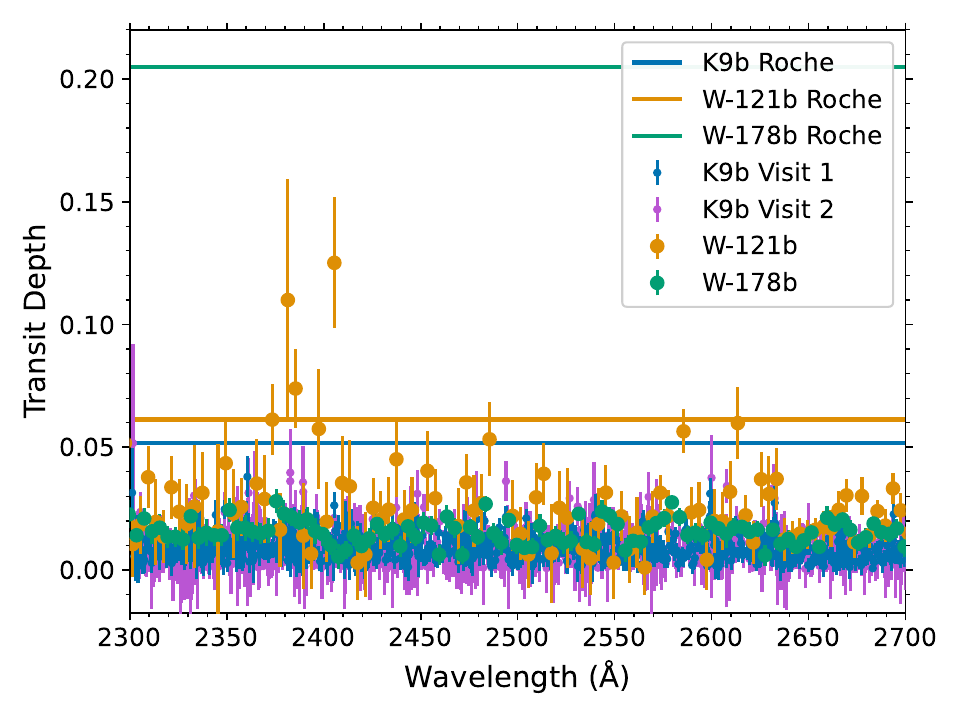}
\includegraphics[width=0.48\textwidth]{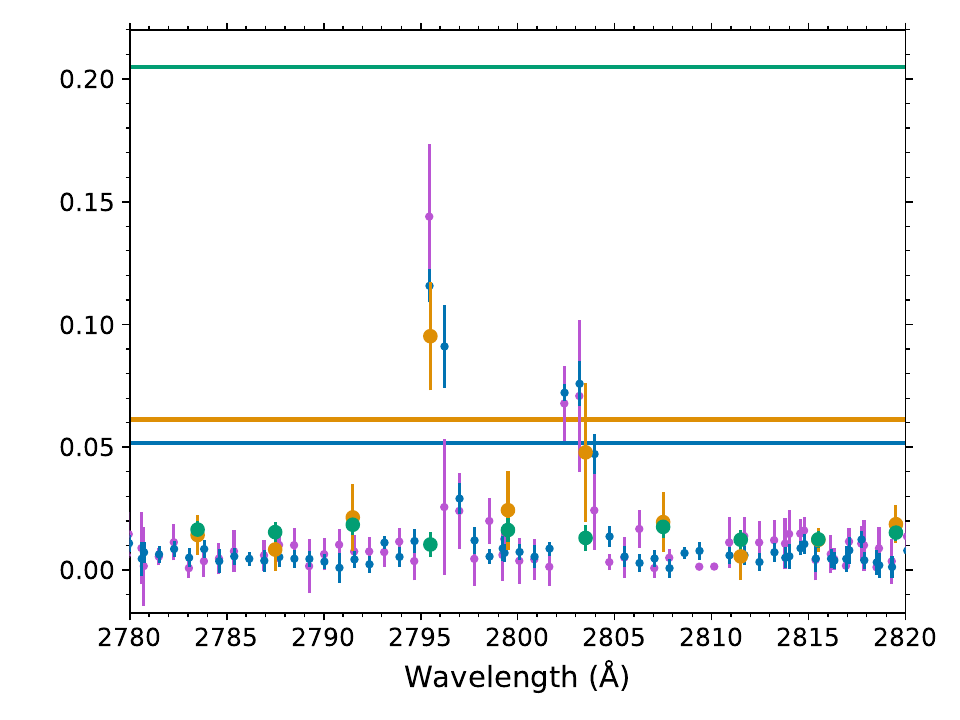}
\caption{A comparison between STIS E230M transit observations of KELT-9b (blue, purple), WASP-121b \citep[gold,][]{sing:2019}, and WASP-178b \citep[green,][]{lothringer:2022}.
\label{fig:stis_compare}}
\end{figure*}

\begin{figure*}
    \centering
    \includegraphics[width=1\linewidth]{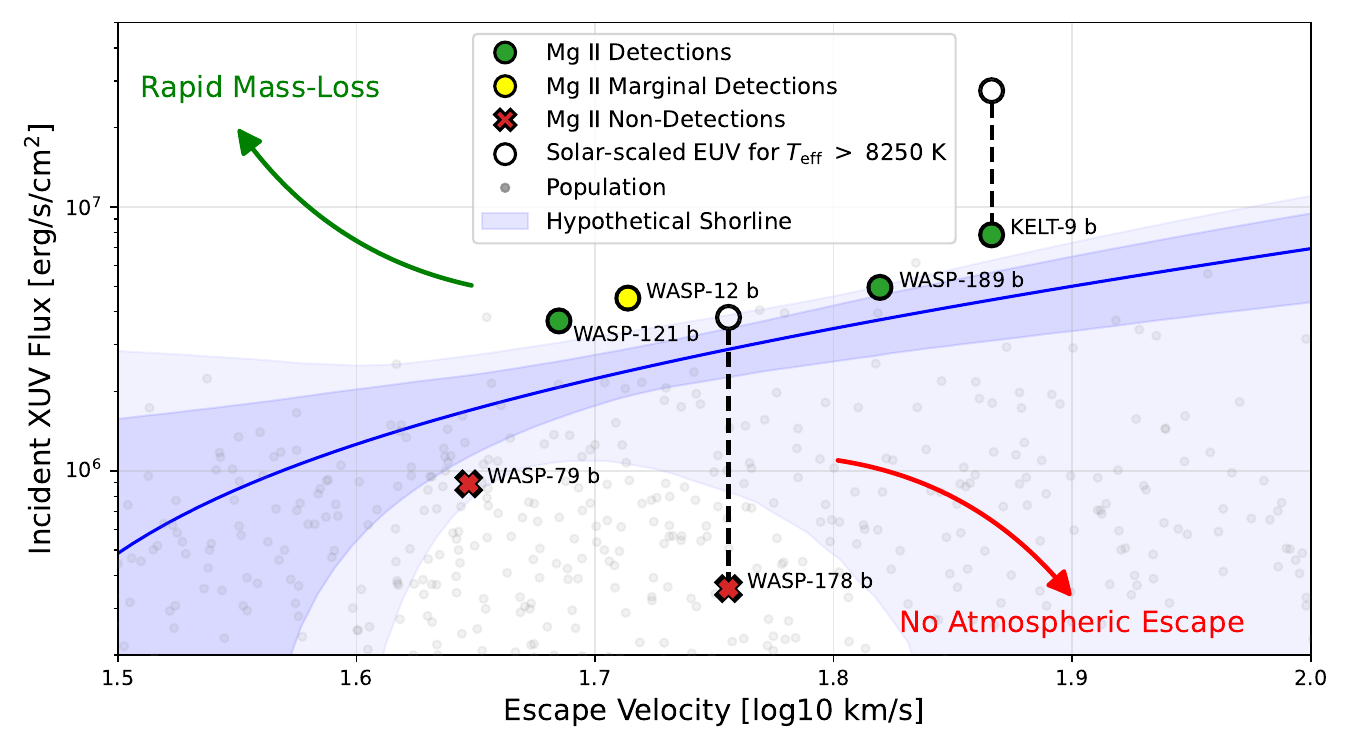}
    \caption{The collection of exoplanets with measurements of Mg II in the NUV. The XUV irradiation (90-912$\rm\AA$) each planet is plotted versus the planet's escape velocity. Small grey points show the background population of known planets. We then run a logistical regression to constrain the median location of the ``shoreline" (blue line) and the 1- and 2-$\sigma$ confidence regions (blue shaded region). We assumed a scaled solar spectrum to calculate the XUV irradiation, but for planets around stars where $T_{\mathrm{eff}} >$ 8250~K, we show the XUV flux assuming a blackbody at the stellar $T_{\mathrm{eff}}$. While KELT-9b would be expected to show escaping Mg II in either scenario, WASP-178b's non-detection can be explained through a lack of XUV irradiation due to $T_{\mathrm{eff}} >$ 8250~K.}
    \label{fig:shoreline}
\end{figure*}

\subsection{Comparison to Previous NUV Observations: The Mg II Cosmic Shoreline} \label{sec:disc:stis}

As mentioned in Section~\ref{sec:intro}, two other well-studied ultra-hot Jupiters, WASP-121b and WASP-178b, have been observed with STIS/E230M \citep{sing:2019,lothringer:2022} and provide an interesting comparison to our results here with KELT-9b.

KELT-9b and WASP-121b happen to share very similar Roche lobe sizes; KELT-9b is more than twice the mass of WASP-121b, but orbits a star that is more than twice the mass of WASP-121. WASP-121b was found to have both strong Fe II and Mg II escape as shown in Figure~\ref{fig:stis_compare}. While KELT-9b shows a very similar Mg~II transit depth, WASP-121b clearly has stronger Fe II escape.

On the other hand, WASP-178b has a much larger Roche radius than both WASP-121b and KELT-9b, thanks to its larger orbital radius. If either Fe II or Mg II were filling WASP-178b's Roche lobe, then the absorption would be expected to be more than twice the depth of those seen in WASP-121b or KELT-9b. Curiously, however, absolutely no absorption is seen in WASP-178b's high-resolution NUV transit spectrum, suggesting some fundamental difference between WASP-178b and the other two planets. We confirm the absence of escaping Mg II in WASP-178b as first presented in \cite{lothringer:2022} with our custom pipeline presented here. While systematic noise makes the identification of continuum-level features difficult, no analysis of the data shows any excess absorption associated with Mg II, let alone the several-percent transit depths that would indicate escape \citep{lothringer:2022,fossati:2025}. 

One hypothesis for why WASP-178b does not show escaping Mg II is because the planet orbits a star that is hotter than 8250~K. Above this temperature, stars appear to have especially weak X-ray and extreme ultra-violet (XUV) radiation, likely related to the lack of surface convection in fully-radiative stars \citep{fossati:2018b}. This XUV irradiation is thought to be responsible for driving hydrodynamic atmospheric escape in the energy-limited regime \citep{salz:2016}. Thus, WASP-178b experiences relatively less XUV irradiation than planets at similar equilibrium temperatures and therefore does not experience sufficient XUV irradiation to drive the escape of Mg II.

Interestingly, KELT-9b is one of only two other planets around a host star hotter than 8250~K, with the other one being KELT-20b/MASCARA-2b \citep{lund:2017,talens:2018}. However, KELT-9b is so highly irradiated, that even with the reduced XUV irradation from a star hotter than 8250~K, there is still enough XUV irradiation to drive escape of Mg II. 

This can be seen if one considers Mg II in the context of a ``Cosmic Shoreline" \citep{zahnle:2017}, now widely used as a hypothesis for atmosphere retention around small exoplanets \citep[e.g.,][]{moran:2023,chatterjee:2024}. Figure~\ref{fig:shoreline} compares the incident XUV flux to the planetary escape velocity, plotting planets with measurements of Mg II. Planets experiencing high levels of XUV irradiation and with sufficiently low escape velocities are expected to undergo the most mass-loss via hydrodynamic atmospheric escape. Planets with low escape velocities (i.e., high surface gravity) or with low XUV irradiation would be expected to show more stable, non-escaping atmospheres. The XUV irradiation is calculated using a solar XUV spectrum between 89 and 912\AA{} from \cite{fossati:2018b} scaled by the stellar radius, stellar effective temperature, and the planet's semi-major axis. For stars hotter than 8250~K, the XUV irradiation is calculated assuming a blackbody at the effective temperature, simulating a lack of chormospheric activity. Further modeling of the Cosmic Shoreline could improve on this with more bespoke chromospheric models.

The $>3\sigma$ detections of escaping Mg II in WASP-121b \citep{sing:2019}, WASP-189b \citep{sreejith:2023}, KELT-9b (this work), and, tentatively, WASP-12b \citep[2.8$\sigma$,][]{fossati:2010} clearly lie above those from the non-detections of WASP-79b \citep{gressier:2023} and WASP-178b \citep{lothringer:2022}, \emph{but only if WASP-178b experiences a reduced XUV flux from its host star as hypothesized in \cite{fossati:2018b}.} If WASP-178b experienced a similar amount of XUV irradiation as other planets like WASP-121b, then we would expect Mg II to be clearly escaping from the planet.

To quantify the location of the hypothetical Mg II Cosmic Shoreline, we use a logistical regression to calculate the probability of a detection as a function of escape velocity and incident XUV flux, treating WASP-121b, WASP-189b, KELT-9b, and WASP-12b as detections (i.e., 1), and WASP-79b and WASP-178b as non-detections (i.e., 0). The probability is modeled as 

\begin{equation}
    p = \sigma(\beta_0 +\beta_1 V_\mathrm{esc}+ \beta_2 F_\mathrm{XUV} ),
\end{equation}

\noindent where $\sigma(x)$ is the logistic sigmoid function

\begin{equation}
    \sigma(x) = \frac{1}{1+e^{-x}},
\end{equation}

\noindent which effectively converts any value $x$ to be between 0 and 1. We find $\beta_0 = 392^{+617}_{-385}$, $\beta_1 = -15.8^{+10.8}_{-13.7}$, and $\beta_2 = 1.2^{+0.24}_{-0.40} \times 10^{-4}$. While the effect of escape velocity on Mg II escape can be expected from first principles, as well as from the Solar System's own Cosmic Shoreline \citep{zahnle:2017}, the current data show that the XUV flux appears to be the most significant factor in determining whether Mg II is escaping or not.

More observations of Mg II for planets orbiting a variety of host stars are necessary to clearly establish the hypothesis, but these results show that atmospheric escape from hot and ultra-hot Jupiters provide unique methods with which we can test the ``Cosmic Shoreline" framework in a level of detail that is currently difficult for terrestrial planets. Future work may explore the role that magnetic fields (both planetary and stellar) and surface gravity may have on the degree to which Mg II is seen escaping.

\section{Conclusion}\label{sec:conc}

We analyzed KELT-9b's HST/STIS E230M transmission spectrum, detecting escaping Mg II and Fe II features, and showing their mass loss rates and line width. We measured the presence of these lines, as well as the extent of these ions out to $R_p^2/R_s^2 = 0.15$. These lines exceed KELT-9b's Roche lobe at approximately $R_p^2/R_s^2 = 0.05$, and imply that gas is escaping the planet's gravitational influence.

We found similar mass loss rates to \cite{yan:2018} and \cite{wyttenbach:2020}, but by contrast, found that nearly double the broadening velocity was needed to fit our model to the Mg II doublet, which suggest some hydrodynamic differences between escaping clouds of hydrogen and Mg II clouds. These results highlight the usefulness of Mg II as tracer of atmospheric escape, which can be readily detected a relatively high SNR around planets orbiting UV-bright host stars.

We compared our observations of KELT-9b to other planets with Mg II measurements and found evidence for a Mg II ``Cosmic Shoreline", where planets with sufficiently high XUV irradiation and low planetary escape velocities show Mg II escape, while planets experiencing less XUV irradiation do not show Mg II escape. In this context, WASP-178b's lack of atmospheric escape \citep{lothringer:2022} can be explained by orbiting a star without chromospheric activity, consistent with the hypothesis from \cite{fossati:2018b}. The KELT-9 system is thought to be in a similar regime, but the planet orbits so close to its even hotter host star that detectable mass-loss rates are still possible. This result highlights the importance of XUV irradiation in driving atmospheric mass-loss.

Full phase coverage of the transit, especially along its ingress and egress, would help determining the shape of the escaping gas, especially the length of its tail. Additionally, more nuanced, MHD and/or multi-dimensional atmospheric models for the planet could help explain the observed escape rate, Doppler shifts, and line broadening. Future observations of other planets along the Mg II ``Cosmic Shoreline" will also help elucidate the behavior of hydrodynamic atmospheric escape.

\section{Acknowledgments}

We thank the anonymous referee for their helpful comments and suggestions that improved the manuscript. We thank Fei Yan for sharing the H$\alpha$ observations from \cite{yan:2018}. We thank Dion Linnson for help with \texttt{sunbather}. We thank Brett Morris for pointing out that our best-fit mass-loss rate of $\dot{M} = 10^{12.5}$ g/s is approximately twice the mass of all elephants on Earth per second. This research is based on observations made with the NASA/ESA Hubble Space Telescope obtained from the Space Telescope Science Institute, which is operated by the Association of Universities for Research in Astronomy, Inc., under NASA contract NAS 5–26555. These observations are associated with program(s) HST-GO-16270. Support for program HST-GO-16270 was provided by NASA through a grant from the Space Telescope Science Institute, which is operated by the Association of Universities for Research in Astronomy, Inc., under NASA contract NAS 5–26555. All the {\it HST} data used in this paper can be found in MAST: \dataset[10.17909/sfym-p902]{http://dx.doi.org/10.17909/sfym-p902}. 

%

\vspace{5mm}
\facilities{HST(STIS)}


\software{
          \texttt{astropy} \citep{astropy:2018},  
          \texttt{PHOENIX} \citep{hauschildt:1999}, 
          \texttt{sunbather} \citep{linssen:2024},
          \texttt{NumPy} \citep{numpy:2020},
          \texttt{SciPy} \citep{virtanen:2020},
          \texttt{batman} \citep{kreidberg:2015:batman}
          }



\clearpage

\bibliography{ms_new}{}
\bibliographystyle{aasjournal}



\end{document}